\documentclass[a4paper,11pt]{article}
\pdfoutput=1 

\usepackage{jinstpub} 

\usepackage[]{subfigure}
 
\usepackage{graphicx}
\usepackage{stackengine}
\usepackage{upgreek}
\usepackage{eqnarray,amsmath}
\usepackage{floatrow}

\newcommand{\nucl}[2]{$\,^{#2}$#1\xspace}

\usepackage{xspace}

\def \ds20 {DarkSide-20k\xspace}
\def \g4ds{G4DS\xspace}
\def \ar39 {$\,^{39}$Ar\xspace}
\def \f90 {f$_{90}$\xspace}
\def \ar {\nucl{Ar}{39}\xspace}

\newcommand\mycaption[2]{\centering \caption{#1\xspace #2}}
\usepackage{booktabs} 
\usepackage{multirow}
\usepackage{array}

\usepackage{marginnote}

\newfloatcommand{capbtabbox}{table}[][\FBwidth]

\title{Simulation of argon response and light detection in the DarkSide-50 dual phase TPC}

\author[a]{P.~Agnes}\emailAdd{pagnes@in2p3.fr} 
\author[b]{I.~F.~M.~Albuquerque}
\author[c]{T.~Alexander}
\author[d]{A.~K.~Alton}
\author[c]{D.~M.~Asner}
\author[c]{H.~O.~Back}
\author[e]{K.~Biery}
\author[f]{V.~Bocci}
\author[g]{G.~Bonfini}
\author[h]{W.~Bonivento}
\author[i,g]{M.~Bossa}
\author[j,k]{B.~Bottino}
\author[l,m]{F.~Budano}
\author[l,m]{S.~Bussino}
\author[n,h]{M.~Cadeddu}
\author[n,h]{M.~Cadoni}
\author[o]{F.~Calaprice}
\author[a,g]{N.~Canci}
\author[g]{A.~Candela}
\author[n,h]{M.~Caravati}
\author[k]{M.~Cariello}
\author[g]{M.~Carlini}
\author[p,q]{S.~Catalanotti}
\author[p,q]{V.~Cataudella}
\author[r,g]{P.~Cavalcante}
\author[s]{A.~Chepurnov}
\author[h]{C.~Cical\`o}
\author[q]{A.~G.~Cocco}
\author[p,q]{G.~Covone}
\author[t,u]{D.~D'Angelo}
\author[g]{M.~D'Incecco}
\author[k,i]{S.~Davini}
\author[p,q]{A.~de~Candia}
\author[v]{S.~De~Cecco}
\author[g]{M.~De~Deo}
\author[p,q]{G.~De~Filippis}
\author[l,m]{M.~De~Vincenzi}
\author[w]{A.~V.~Derbin}
\author[p,q]{G.~De~Rosa}
\author[n,h]{A.~Devoto}
\author[o]{F.~Di~Eusanio}
\author[g,u]{G.~Di~Pietro}
\author[f,v]{C.~Dionisi}
\author[x]{E.~Edkins}
\author[a]{A.~Empl}
\author[y]{A.~Fan}
\author[p,q]{G.~Fiorillo}
\author[z]{K.~Fomenko}
\author[aa]{D.~Franco}\emailAdd{dfranco@in2p3.fr}
\author[g]{F.~Gabriele}
\author[o,u]{C.~Galbiati}
\author[f,v]{S.~Giagu}
\author[ab]{C.~Giganti}
\author[o]{G.~K.~Giovanetti}
\author[g]{A.~M.~Goretti}
\author[ac]{F.~Granato}
\author[s]{M.~Gromov}
\author[ad]{M.~Guan}
\author[e]{Y.~Guardincerri}
% ...

\affiliation[a]{Department of Physics, University of Houston, Houston, TX 77204, USA}
\affiliation[b]{Instituto de F\'isica, Universidade de S\~ao Paulo, S\~ao Paulo 05508-090, Brazil}
\affiliation[c]{Pacific Northwest National Laboratory, Richland, WA 99352, USA}
\affiliation[d]{Physics Department, Augustana University, Sioux Falls, SD 57197, USA}
\affiliation[e]{Fermi National Accelerator Laboratory, Batavia, IL 60510, USA}
\affiliation[f]{INFN Sezione di Roma, Roma 00185, Italy}
\affiliation[g]{INFN Laboratori Nazionali del Gran Sasso, Assergi (AQ) 67100, Italy}
\affiliation[h]{INFN Cagliari, Cagliari 09042, Italy}
\affiliation[i]{Gran Sasso Science Institute, L'Aquila 67100, Italy}
\affiliation[j]{Physics Department, Universit\`a degli Studi di Genova, Genova 16146, Italy}
\affiliation[k]{INFN Genova, Genova 16146, Italy}
\affiliation[l]{INFN Roma Tre, Roma 00146, Italy}
\affiliation[m]{Mathematics and Physics Department, Universit\`a degli Studi Roma Tre, Roma 00146, Italy}
\affiliation[n]{Physics Department, Universit\`a degli Studi di Cagliari, Cagliari 09042, Italy}
\affiliation[o]{Physics Department, Princeton University, Princeton, NJ 08544, USA}
\affiliation[p]{Physics Department, Universit\`a degli Studi ``Federico II'' di Napoli, Napoli 80126, Italy}
\affiliation[q]{INFN Napoli, Napoli 80126, Italy}
\affiliation[r]{Virginia Tech, Blacksburg, VA 24061, USA}
\affiliation[s]{Skobeltsyn Institute of Nuclear Physics, Lomonosov Moscow State University, Moscow 119991, Russia}
\affiliation[t]{Physics Department, Universit\`a degli Studi di Milano, Milano 20133, Italy}
\affiliation[u]{INFN Milano, Milano 20133, Italy}
\affiliation[v]{Physics Department, Sapienza Universit\`a di Roma, Roma 00185, Italy}
\affiliation[w]{Saint Petersburg Nuclear Physics Institute, Gatchina 188350, Russia}
\affiliation[x]{Department of Physics and Astronomy, University of Hawai'i, Honolulu, HI 96822, USA}
\affiliation[y]{Physics and Astronomy Department, University of California, Los Angeles, CA 90095, USA}
\affiliation[z]{Joint Institute for Nuclear Research, Dubna 141980, Russia}
\affiliation[aa]{APC, Universit\'e Paris Diderot, CNRS/IN2P3, CEA/Irfu, Obs de Paris, USPC, Paris 75205, France}
\affiliation[ab]{LPNHE, Universit\'e Pierre et Marie Curie, CNRS/IN2P3, Sorbonne Universit\'es, Paris 75252, France}
\affiliation[ac]{Physics Department, Temple University, Philadelphia, PA 19122, USA}
\affiliation[ad]{Institute of High Energy Physics, Beijing 100049, China}
% ...

\abstract{
A Geant4-based Monte Carlo package named G4DS has been developed to simulate the response of DarkSide-50, an experiment  operating since 2013 at LNGS, designed to detect WIMP interactions in liquid argon. In the process of WIMP searches, DarkSide-50 has   achieved two  fundamental milestones: the rejection of   electron recoil background with a power of $\sim$10$^7$, using  the pulse shape discrimination technique, and the measurement of the residual $^{39}$Ar contamination in underground argon,  $\sim$3 orders of magnitude lower with respect to atmospheric argon. These results  rely on the accurate simulation of the detector response to the liquid argon scintillation, its ionization, and  electron-ion recombination processes.  This work provides a complete overview of the DarkSide Monte Carlo and of its performance, with a particular focus on PARIS, the custom-made liquid argon response model.

}

\begin{document} 
\sloppy
\maketitle

\section{Introduction}
\label{sec:intro}
Evidence from several astronomical observations and calculations shows gravitational patterns in the universe which require interactions between baryonic matter and a different form of non-luminous matter. This ``dark matter'' is  a new form of matter that interacts gravitationally, but not via electromagnetic or strong forces, with baryons. A candidate for this dark matter is the Weakly Interacting Massive Particle, WIMP, which would exist in all gravitationally bound clusters of matter. Many searches for the existence of WIMPs are currently being performed.  
 
Experiments looking for the direct detection of dark matter attempt to observe the recoil of an atomic nucleus which has undergone a weak interaction with a WIMP.  A  large number of direct dark matter searches and related R\&D efforts are currently in progress, which demonstrate the present importance   of this topic. Several detector types using different techniques are either acquiring data or under development in various   underground laboratories.  In order to identify signals which  are due to non-standard, extremely rare interactions, these detectors all   share the common requirements of sensitivity to the very low-energy deposition occurring in a WIMP-nuclear collision and the ability to discriminate against common,  naturally occurring backgrounds.  

 The best WIMP-nuclear cross section limits for WIMP masses larger than  10~GeV are achieved with  dual-phase, noble--liquid time projection chambers (TPC)  \cite{Akerib:2016vxi,Yang:2016odq, Aprile:2017iyp}.  Noble  liquids are attractive targets for rare low energy particle interactions because of their potential radio-purity, their high density,  and their response to various particle interactions. In a dual--phase TPC,  scintillation photons and ionization electrons, emitted during a WIMP-nucleus interaction, can both be  detected with high resolution. The scintillation light (S1) from  this interaction is  detected by  photosensors, while the ionization electrons drift under the influence of an electric field into a gas phase at the top of the detector where they are extracted. The extracted electrons  produce a secondary pulse of light (S2) by electroluminescence in the  gas. The S1 and S2 pulses provide  all the necessary information needed to reconstruct the position of the  interaction vertex, and aid the identification and discrimination  of background events.  
 
Liquid argon (LAr) excitation occurs with two dimer states having  lifetimes differing by 2-3 orders of magnitude.  The relative ratio between the two dimers is governed by the density of the energy loss due to the ionizing particle.  Therefore the ratio differs for excitations due to electron as opposed to nuclear recoils, and thus the S1 pulse shape is  determined by the nature of the interacting particle.  This provides extraordinary discrimination between nuclear (WIMP-induced), and electron recoils. 

DarkSide-50 uses a 46~kg active--mass, dual--phase argon TPC which is installed at LNGS and has been in operation since 2013.  It has recently confirmed the exceptional pulse shape discrimination (PSD) of LAr \cite{Agnes:2014bvk} by identifying and rejecting approximately 1.5$\times$10$^{7}$ electron recoil events, mainly due to $^{39}$Ar decays.  The $^{39}$Ar component is  a cosmogenically produced background which is naturally present in argon extracted from the atmosphere.  Despite the large PSD power, the intrinsic $^{39}$Ar contamination ($\sim$1 Bq/kg with 269 yr half-life) is so high that it prevents using atmospheric argon in a large volume experiment. The DarkSide collaboration recently overcame this problem by extracting argon from deep underground~\cite{Agnes:2015ftt}, which is naturally shielded from cosmic rays and hence depleted in $^{39}$Ar. DarkSide-50, which was filled with underground   argon in April 2015, measured a  $^{39}$Ar depletion factor  of $\sim$1400 ~\cite{Agnes:2015ftt}. This result is fundamental to the proposed DarkSide-20k (DS-20k) detector, which will use some 20 tonne of Ar depleted in $^{39}$Ar. The operation of DS-20k is foreseen to start in 2020.  


The success of DarkSide-50   strongly relies on a full understanding of the detector response, and this is possible thanks to detailed and accurate descriptions of the geometry,  material properties, and physical processes as simulated in a Geant4--based DarkSide Monte Carlo simulation toolkit, G4DS (GEANT4.10.0.0 \cite{Agostinelli2003250}). G4DS plays a pivotal role in the DarkSide-50 analysis, and reproduces  data with high accuracy.  This paper  provides a full description of G4DS and of the custom physics process developed to describe the physics of the S1 and S2 signals by calibrating it with DarkSide-50 data.




\section{The DarkSide-50 detector}
\label{sec:ds50}
\begin{figure}[h]
\centering
\includegraphics[width=.45\textwidth]{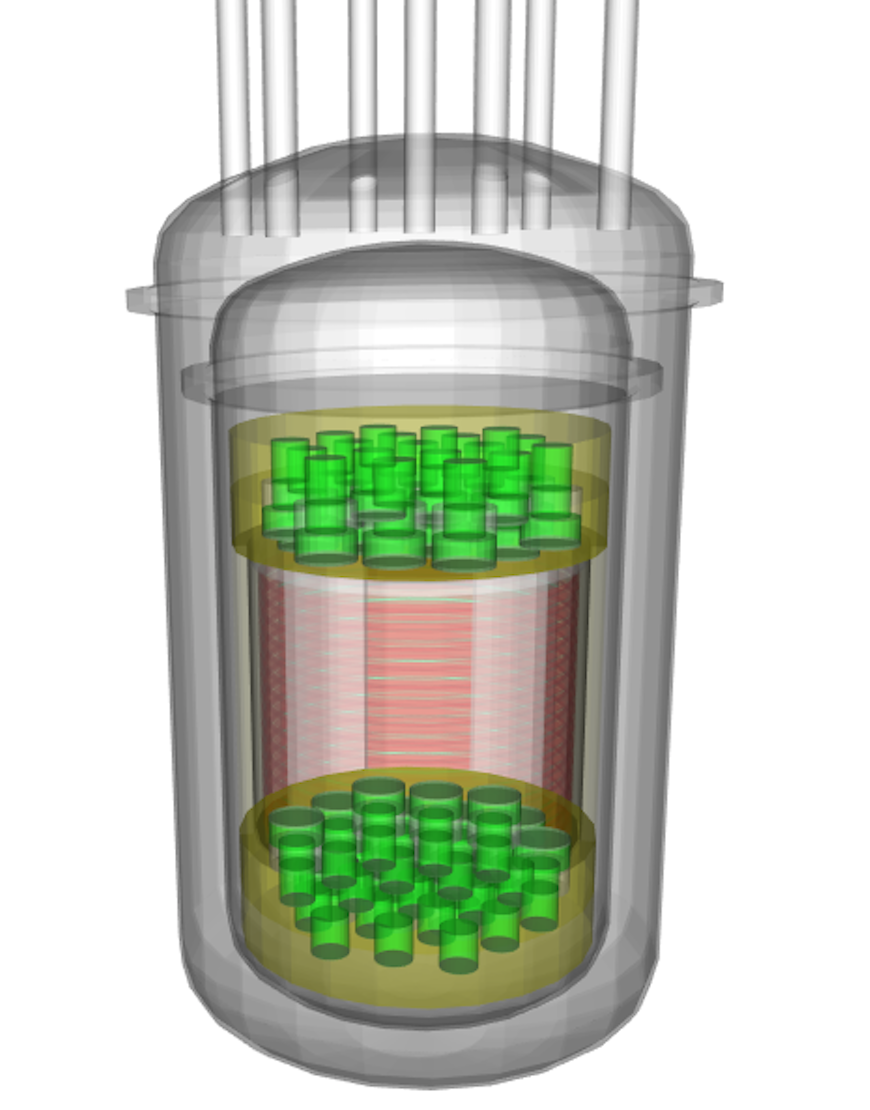}
\caption{\label{fig:tpcds50}Simulated TPC geometry.}
\end{figure}

As shown in Figure \ref{fig:tpcds50}, the  DarkSide-50 target mass of LAr  (46.4$\pm$0.7~kg) is  contained in  a cylindrical region, laterally confined by a 2.54 cm--thick PTFE reflector, and viewed by a two arrays of 19 low-background Hamamatsu R11065 3'' photomultiplier tubes (PMTs) at the top and bottom of the TPC. The  quantum efficiency of the PMTs is 34~\% at 420 nm, and they are entirely immersed in  LAr.  They view the TPC interior through fused--silica windows, which are coated on both faces by 15~nm thick  conductive indium tin oxide (ITO) films. The ITO  serves as a grounded anode at the top and a HV cathode at the  bottom of the TPC. The windows and the radial-defining PTFE  are coated with tetraphenyl butadiene (TPB), which converts ``128'' nm Ar scintillation photons into 420 nm photons, so that signals are detectable by the PMTs. The electron drift system consists of the ITO cathode and anode planes, a radial field cage, and a grid used to extract drifting electrons into the 1~cm gaseous region above the LAr. The grid lies 5~mm below the liquid surface and is a hexagonal mesh which is etched from a 50~$\mu$m-thick stainless steel foil and has an optical transparency of 95\% at normal incidence. The field cage lies outside the cylindrical PTFE wall and consists of copper rings held at a graded potential. It is designed to keep the drift field a uniform 200 V/cm throughout the active volume.   

\begin{figure}[h]
\centering
\includegraphics[width=.66\textwidth]{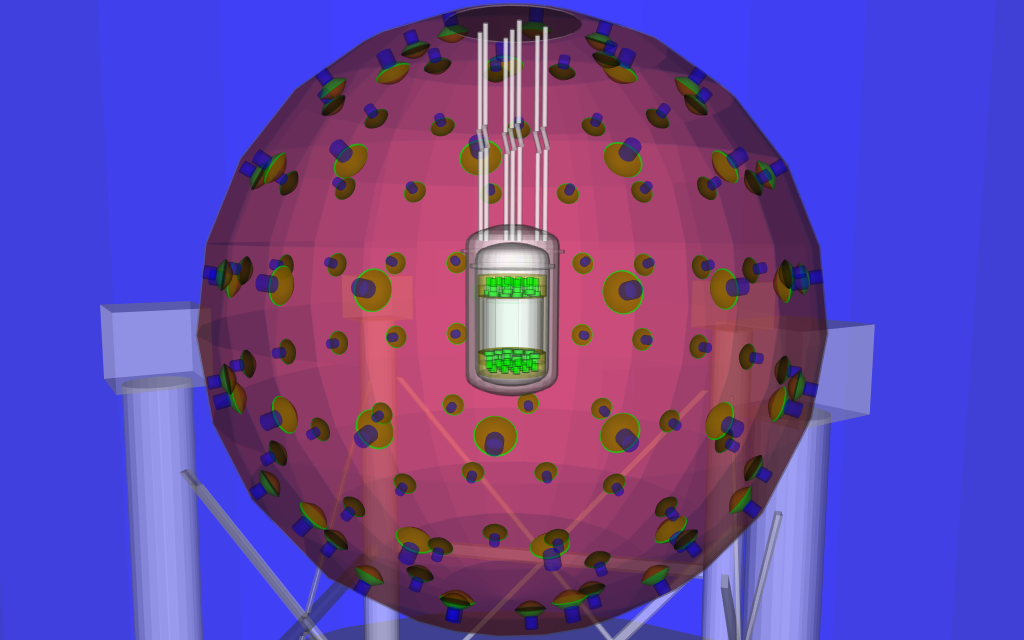}
\caption{\label{fig:ds50}Simulated geometry of the TPC inside the LSV.}
\end{figure}

There are two independent, active veto shields enclosing the TPC.  The outer shield is  1,000--tonne water Cherenkov muon veto with an inner, 30--tonne liquid scintillator  neutron veto, as shown in Figure \ref{fig:ds50}.  G4DS provides a full simulation of  the two vetoes, however details are omitted here since the scope of this work is limited to the simulation of the TPC. More details on the LAr TPC, the liquid scintillator veto, and the water Cherenkov veto can be found in references~\cite{Agnes:2014bvk,Agnes:2015qyz,Agnes:2016nis}.  

\section{The Monte Carlo simulation}

\label{sec:mcsim}
\subsection{G4DS - Geant4-based DarkSide Monte Carlo simulation toolkit}

G4DS  is designed with a modular architecture in order to include a full description of  all the detectors belonging to  the DarkSide program.  These detectors are DarkSide-10~\cite{Alexander201344}, DarkSide-50~\cite{Agnes:2014bvk},  DarkSide-20k~\cite{DS20k, DS20kTDR},  and  ARGO~\cite{1475-7516-2016-08-017}. The last step in this experimental program will have several hundred tonnes of fiducial mass and will be able to reach the  detection sensitivity required to observe solar neutrino signals,  the so--called "neutrino floor".

This paper is focused on the performance of G4DS for DarkSide-50, where it plays a fundamental role in the definition of the data selection criteria,  estimation of  associated efficiencies,    optimization of  reconstruction algorithms, and  evaluation  of  residual backgrounds.   G4DS embeds a rich set of particle generators, detailed  geometry descriptions, properly tuned physical processes, and the full optical propagation of photons produced by scintillation in liquid argon and electroluminescence in gaseous argon.  It  tracks all the generated photons until they reach the active region of the photosensors, where they are converted, after  stochastically surviving  a quantum efficiency cut, into  photoelectrons. The conversion of a  photoelectron into a charge signal is then handled by  electronic simulation, as described in Subsection~\ref{sec:electronics}.  

A large choice of event generators allows simulations  of forbidden beta decays (for example $^{39}$Ar and $^{85}$Kr),  single and chain radioactive decays, cosmic muon and neutron fluxes as measured at LNGS, and AmBe and AmC neutron calibration sources.    The electromagnetic physics list  used in G4DS  is \textit{G4EmLivermore}, whose  lower threshold at 250~eV perfectly matches  the energy range and  accuracy required by DarkSide-50.  The hadronic processes rely on  a custom-made physics list, including the  \textit{High Precision Neutron} (HP) models for  simulations of neutrons with energy lower than 20 MeV. In this range, the most dangerous background is due to neutrons induced by  ($\alpha$, n) and fission reactions. These neutrons are generated with the TALYS~\cite{TALYS}  package.  At higher energies,  an \textit{ad hoc} FLUKA~\cite{Ferrari:2005zk,Fluka:2014} simulation code  was implemented to  study the production rates of secondary particles  and nuclei induced by cosmic ray interactions. G4DS embeds generators to read and track both TALYS and   FLUKA  outputs.

A critical aspect of the DarkSide-50 simulation is in treating the degeneracies that arise because the LAr scintillation response is non-­linear in the presence of an electric field and the light collection efficiency varies over the volume of the detector. The efficiency of light collection  depends on   the optical properties of materials and surfaces in the TPC. These properties have to be tuned based on observables which are independent of particle type and event energy in order to avoid degeneracies with the LAr energy response. Among the variables used for the optical tuning, described in Section~\ref{sec:optics}, are PMT channel occupancies and  asymmetries between the light collection in the  top and bottom arrays of PMT's.  

Once the optical properties are properly constrained, the  DarkSide-50 energy response depends only on the processes responsible for   light emission in liquid and gaseous argon. These processes are modelled in a custom-made Geant4  module  called  Precision Argon Response Ionization and Scintillation (PARIS), which will be described in detail in Section~\ref{sec:paris}. The PARIS module relies on the effective description of the recombination of   electrons with ions, which induces a depletion of  S2  in favor of the S1 signal.

\subsection{Simulation of the electronics and event reconstruction}
\label{sec:electronics}

\begin{figure}[h]
\centering
\includegraphics[width=.8\textwidth]{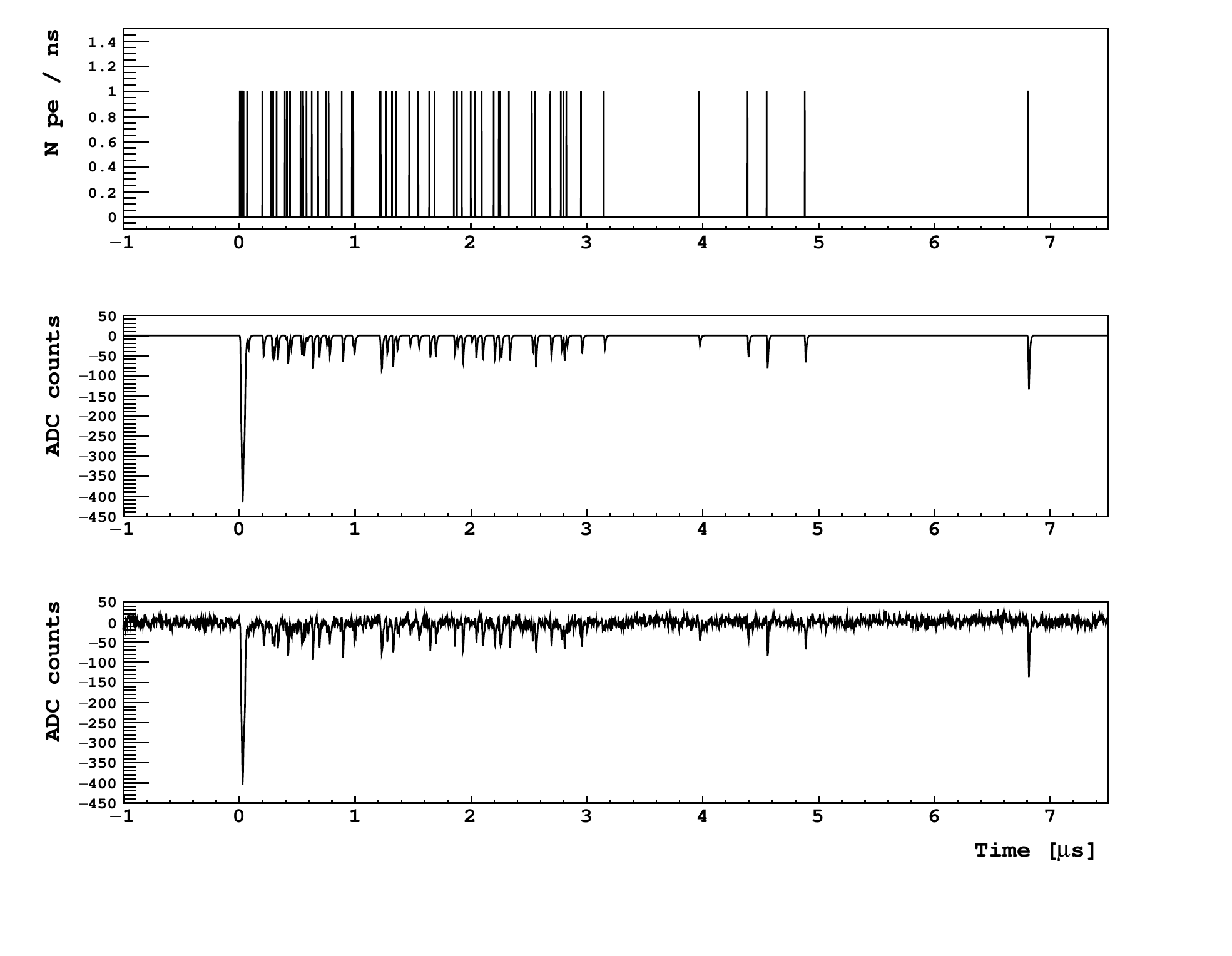}
\caption{Example of a waveform produced by the full Monte Carlo chain.  Top: time distribution of photoelectrons  generated by G4DS; middle:  convolution with the PMT response; bottom: final  waveform including noise extracted from real data.} 
\label{fig:darkart_wf}
\end{figure}

The electronics simulation is developed within \textit{DarkArt}, the \texttt{art}-based framework~\cite{ART} that serves as the  reconstruction code  for DarkSide-50. The \textit{DarkArt} simulation module generates waveforms   for each PMT channel based on the arrival time of the photoelectrons produced by G4DS. Each photoelectron is converted into time and charge, using a single electron response (SER) that is extracted from data. The electronic noise is measured with dedicated  data campaigns using  a random trigger and is directly added to each simulated PMT waveform. An example of a waveform reconstructed with this procedure is shown in figure~\ref{fig:darkart_wf}.

\begin{figure}[h]
\centering
\includegraphics[width=.66\textwidth]{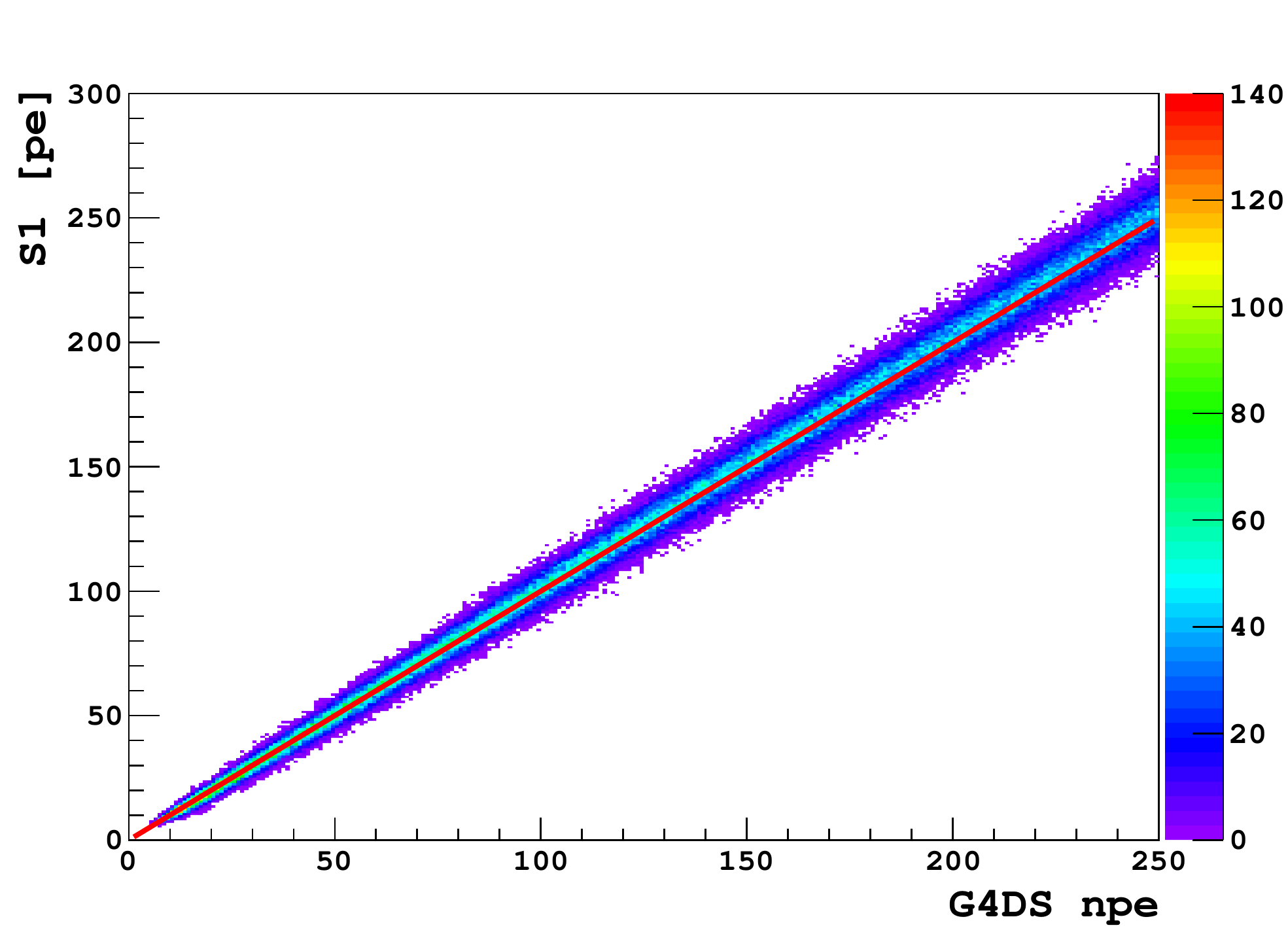}
\caption{Reconstructed S1 as  function of the true number of   photoelectrons generated by G4DS. The deviation from linearity (red line) is estimated at 1.1\%.  }  
\label{fig:darkart_reco}
\end{figure}

The simulated waveforms are then processed by the reconstruction code, which  subtracts the  baseline,  identifies pulses, and integrates the charge.  The difference between the reconstructed S1 variable and  the true  number of photoelectrons generated by G4DS is $\sim$1.1\%,   as shown in figure~\ref{fig:darkart_reco}. The smearing introduced by  the electronics and the reconstruction algorithms is estimated to be 5.3\%, and is negligible compared to the SER.    

\subsection{Fast Mode Approach}

The full simulation of the DarkSide-50 data, including the optical propagation of S1 and S2 photons, electronics simulation, and reconstruction, is CPU and time consuming.  To overcome this problem, several  DarkSide-50 analyses  rely on  a fast reconstruction code based on the G4DS measured variables.  This code stores the visible energy proportional to S2 but does not generate the S2 photons, and bypasses \textit{DarkArt}. It reduces the simulation time by a factor of $\sim$100 and is labelled the  Fast Mode Approach (FMA). The  electronics and reconstruction response for S1 can be simply modelled by smearing the number of photoelectrons with a  Gaussian  resolution, embedding the effects from SER, noise, and signal reconstruction. The main issue with this approach is the identification of multiply scattered events which rely on the reconstruction of multiple S2 pulses. The efficiency of the  DarkSide-50 pulse finder algorithm  \cite{Agnes:2014bvk}  in  distinguishing   overlapping S2 pulses   from multi-sited depositions, depends primarily on the  time difference between the pulses, and hence on  the spatial distance ($\Delta$z) along the drift field between the deposits.

The FMA approach  implements   a \textit{clustering} algorithm,  which defines a  \emph{cluster}  as  a set of energy depositions, whose  positions along the drift field   are too close to be disentangled  as separate S2 pulses. The cluster position is defined as  the  energy-weighted average position of energy depositions. The  cluster energy is the sum of the depositions, with each   independently subject to recombination, as  will be  described in Section \ref{sec:paris}. 

\begin{figure}[h]
\centering
\includegraphics[width=0.6\textwidth]{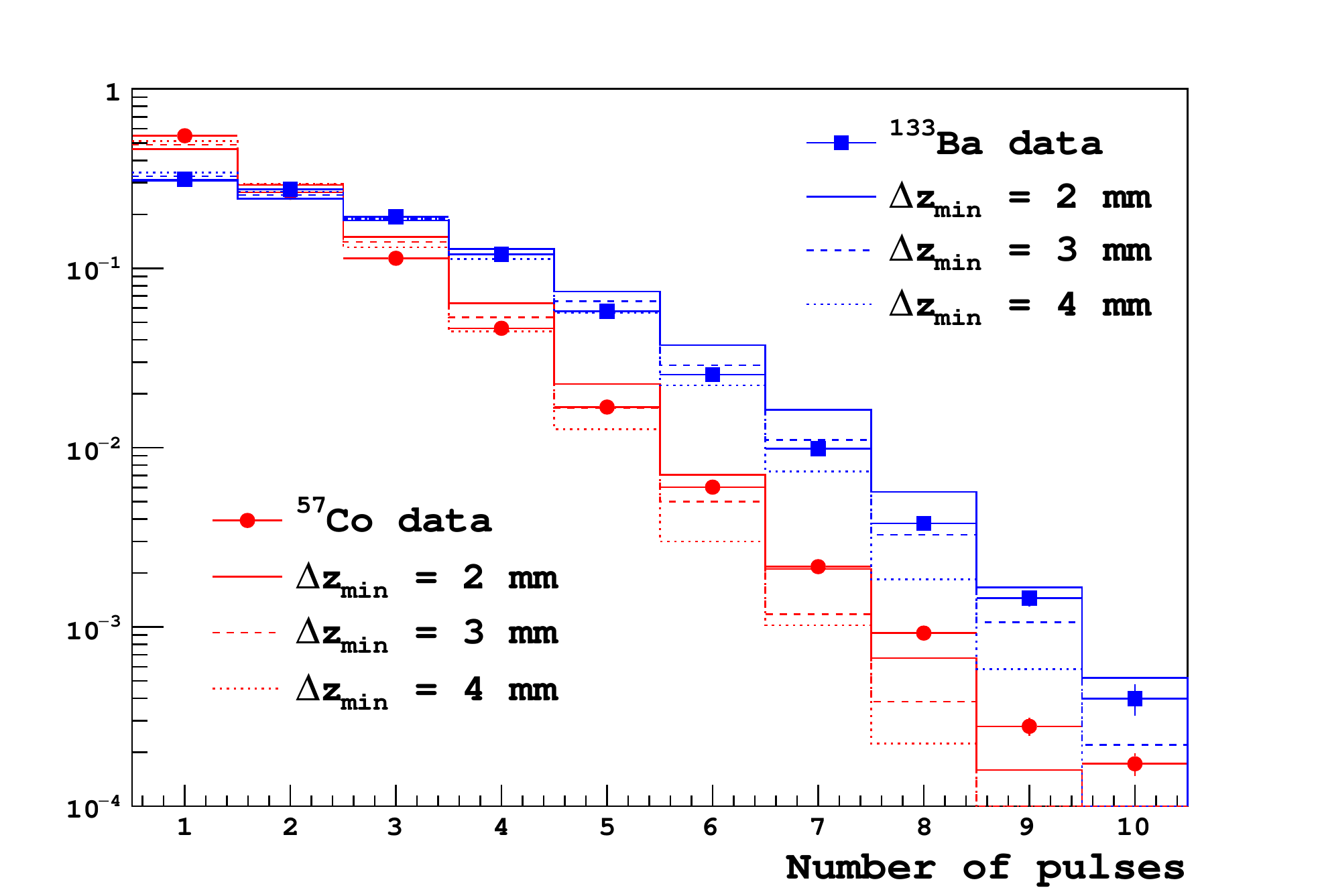}
\caption{Comparison between the number of S2 pulses between   DarkSide-50 calibration data and G4DS using the FMA   reconstruction for different  $\Delta$z$_{min}$ values.} 
\label{fig:clustering}
\end{figure}

The minimum distance ($\Delta$z$_{min}$) between deposits allowing resolution of clusters was tuned using data from  $^{133}$Ba and $^{57}$Co sources, which were located in the liquid scintillator veto close to the cryostat. These two sources are characterized by emission of  356 keV and 122 keV gamma lines, respectively. These reach  the  active volume  after  crossing the cryostat, the TPC wall, and the LAr bath surrounding the TPC. The distributions of S2 pulses identified reconstructing $^{133}$Ba and $^{57}$Co calibration data are compared in figure \ref{fig:clustering} with   simulations processed by the FMA  for different $\Delta$z$_{min}$. 

\begin{figure}[h] 
\centering
\includegraphics[width=1.0\textwidth]{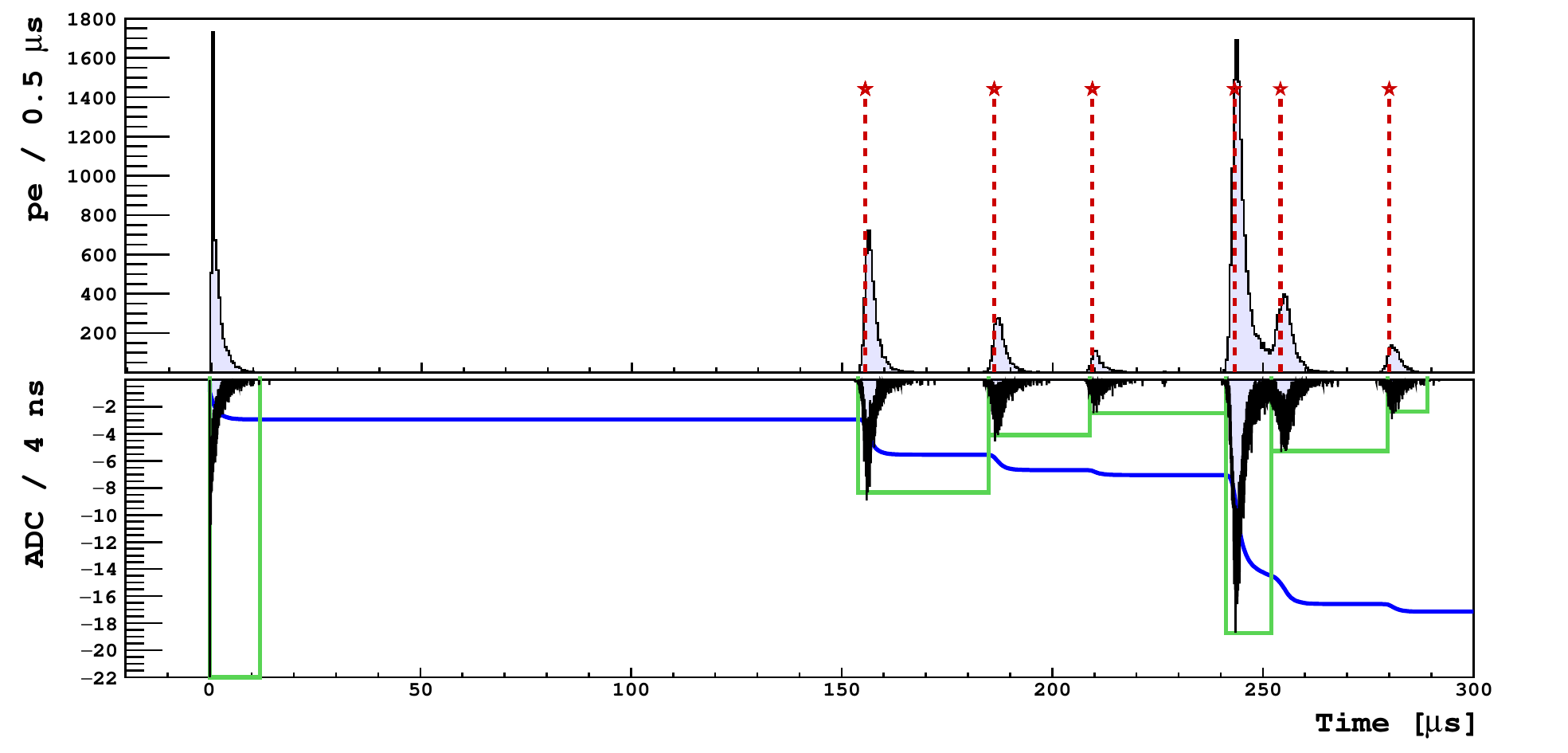}
\caption{Comparison between the G4DS \textit{fast} (top) and   \textit{full} (bottom) reconstruction modes for a single event. TOP:   the S1 pulse (the first), followed by 6 S2 pulses, simulated with G4DS,  are overlapped by  dashed red lines   corresponding to the pulses identified in the fast mode. BOTTOM:  simulated waveform of the event (black), cumulative distribution  (blue line), and the pulses (green boxes) identified by the full  reconstruction mode.  } 
\label{fig:clusters}
\end{figure}

The simultaneous fit of the numbers of S2 pulses of $^{133}$Ba and $^{57}$Co  data, results in $\Delta$z$_{min}$$=$3.5~mm corresponding to a time separation of 3.8~$\mu$s assuming the drift velocity of 0.93~mm/$\mu$s \cite{Agnes:2014bvk}. The FMA clustering algorithm is also able to identify, with good accuracy, the starting time of each pulse, defined as the time of the  first deposit occurring in the cluster. This is  shown in Figure \ref{fig:clusters}, where FMA pulses (dashed red lines) are compared for a single event with those (green boxes) identified by the \textit{DarkArt} full reconstruction approach.


\section{The Liquid Argon TPC optics}
\label{sec:optics}
The propagation of  photons from their generation in liquid or gaseous argon to the photosensor, depends on  a large number of optical properties of the detector materials and their interfaces. The relevant materials affecting the simulation are the liquid and gaseous argon, the fused silica windows, the TPB coatings on each internal surface, the Indium-Tin-Oxide (ITO)  electrodes,  the PTFE reflectors defining the radial edge of the TPC, the grid, and the PMT windows. The sequence of materials crossed by a photon from the center of the TPC to the top, bottom, and radial sides is listed in Table \ref{tab:materials}. 

\begin{table}[h]
\centering
\begin{tabular}{llll}
\hline
\hline
Id & Up         & Down    & Side\\
\hline
1	& Liquid argon			&   Liquid argon  		& Liquid argon \\
2	&Grid				&  TPB 				& TPB 		\\
3	&Liquid argon  			&  ITO 				& PTFE 		\\
4	&Gaseous argon 		& Fused silica 			& 			\\
5	&Condensed liquid argon 	& ITO 				& 			\\
6	&TPB		         	 & Liquid argon 			& 			\\
7	&ITO				 	& PMT cathode or PTFE	& 			\\
8	&Fused silica			& 					& 			\\
9	&ITO				 	& 					& 			\\
10	&Liquid argon  			& 					& 			\\
11	&PMT cathode or PTFE	& 					& 			\\
\hline
\hline
\end{tabular}
\caption{\label{tab:materials}Sequences of materials crossed by a photon from the center toward the top, bottom, and side of the TPC.   }  
\end{table}

Most of these relevant materials are modelled as pure dielectrics, with the exceptions of the ITO coating and the grid, which are described as optical surfaces, and the PTFE which is assumed metallic, i.e a surface which photons cannot penetrate.  The model for the grid, adjusted to optimize CPU simulation time, is an optical surface with a transmittance dependent on the incident photon angle. This is equal to $\sim$95\% for normal incidence.  Non-transmitted photons undergo either reflection or absorption depending on a wavelength dependent function.  

\begin{figure}[h]
\centering
\includegraphics[width=.6\textwidth]{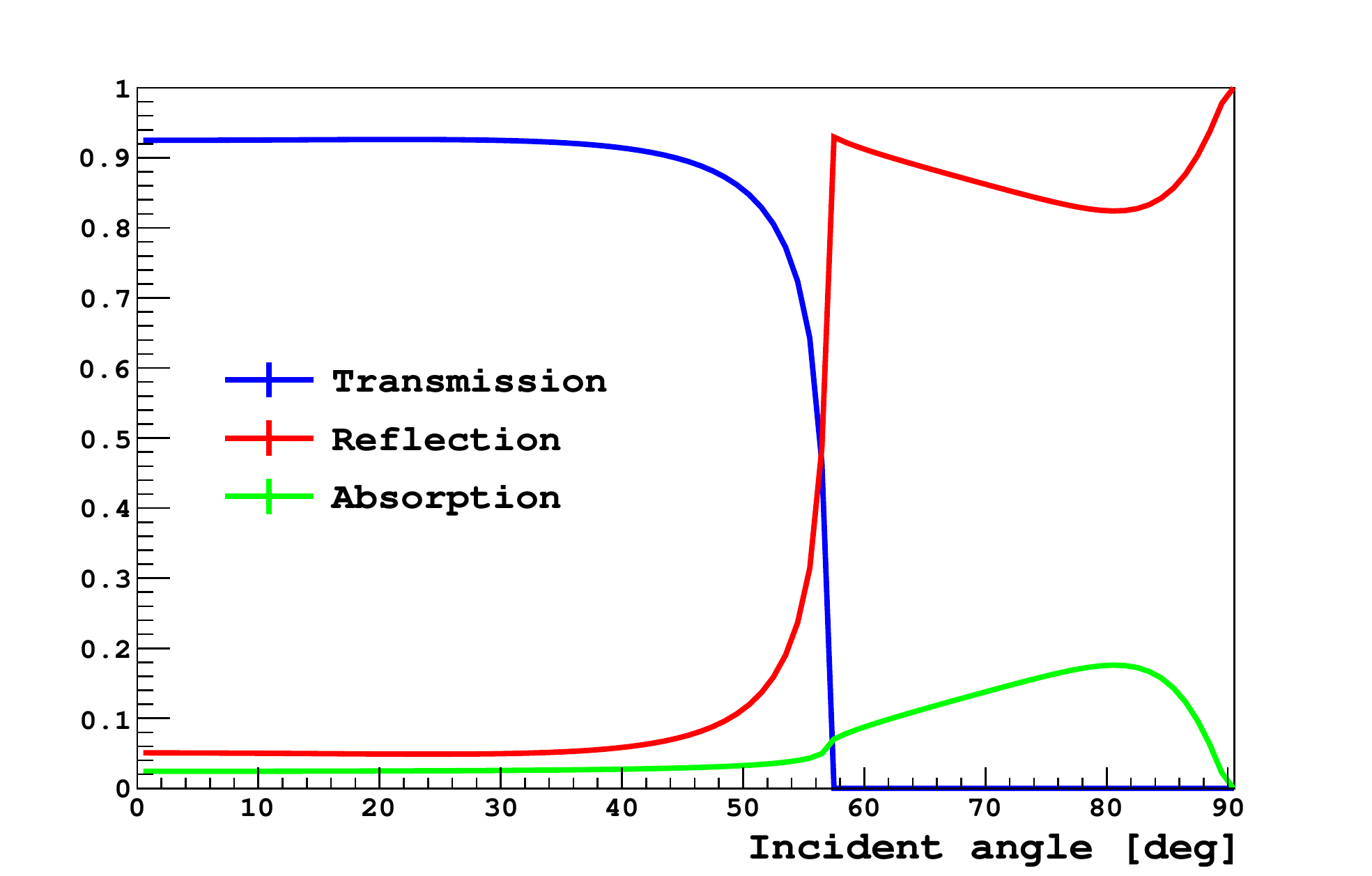}
\caption{ Coefficients of transmission, reflection, and absorption versus  incident angle for ITO, when photons cross, in sequence,   fused silica, ITO, and  liquid argon. } 
\label{fig:ITO}
\end{figure}

The ITO coating has a refractive index with an imaginary (absorptive) component, and requires a custom model. Here, reflectivity is assumed specular, and transmittance and absorbance are described as a function of the wavelength and the material on which the ITO is deposited. Figure \ref{fig:ITO} shows three optical couplings for 420 nm photons as a function of the incident angle for photons on ITO deposited on fused silica in LAr.   

All the internal PTFE radial surfaces are  coated with a TPB wavelength shifter. No reflection coefficients are required for the PMT supports as these are not directly exposed to the UV light  due to limitations imposed by the TPB coated  fused silica  windows. The PTFE is  modelled with a Lambertian reflectivity of 98\%  and 7.5\% for the visible and UV ranges, respectively. Light diffusion from the lateral walls is dominated by TPB re-emission in the visible range, since the probability that a UV photon survives before absorption in the TPB ($\sim$100~$\mu$m thick)  is negligible.  The TPB is assumed to re-emit a single visible photon for each absorbed  UV photon\footnote{V. M. Gehman  \textit{et. al.}~\cite{Gehman2011116}   measured   1.2  visible photons  emitted for each UV (128 nm) photon. This is not considered in this work because  of the low significance of the measurement ($\sigma$$\sim$1.6).} with a characteristic time of 1.5~ns \footnote{E. Segreto~\cite{PhysRevC.91.035503} found delayed light emission components from TPB. As this is a recent result it has not been included in G4DS. However, tuning of the LAr scintillation time profile almost entirely renormalizes the effect of any  TPB delayed re-emission, as shown in Section~\ref{sec:paris}.}.   

The PMT cathodes are modelled as a dielectric, but with an arbitrarily reduced  absorption length, in order to  fully absorb the transmitted visible photons in a few micrometers. Absorbed photons are converted to photo-electrons, if they survive the PMT quantum efficiency cut which is a function of the  wavelength.   

All the  remaining detector materials adopt standard Geant4 modelling, with parameters tuned by the DarkSide-50 data. The tuning procedure requires calibrations of the optical response of the detector before tuning the energy response. This ordering is required to break   degeneracies between non-linearities in the LAr energy response and the non-uniformities in the photon collection efficiency. A key element for optical tuning requires identification of observables which are independent of energy and interacting particles, such as:  

\begin{itemize}
\item photon collection efficiency as a function of the event position (collection efficiency);
\item ratio of light collected on the top and on the bottom PMT arrays as a function of the vertex position in the drift direction (top/bottom ratio);
\item fraction of S1 light observed by each PMT (S1 channel occupancy);
\item fraction of S2 light observed by each PMT (S2 channel occupancy).
\end{itemize}

The reference sources used for the optical tuning are $^{39}$Ar and $^{83m}$Kr.  $^{39}$Ar is a beta emitter with a Q-value = 565 keV and $\tau_{1/2}$ = 269 y. It  is produced by cosmic ray interactions with $^{40}$Ar, and hence is present in  atmospheric argon,  which has an activity of $\sim$1~Bq/kg  \cite{Benetti:2006az}. The atmospheric argon campaign in DarkSide-50 provides beta events that are uniformly distributed in the TPC. The isomeric state of $^{83m}$Kr decays with a $\tau_{1/2}$ = 1.83~h, emitting a  fast ($\tau_{1/2}$ = 156 ns) cascade of two 32.1 and 9.4 keV electrons,  providing  a clear and identifiable peak in S1. Gaseous $^{83m}$Kr was injected directly into the TPC, and decayed  into  stable $^{83}$Kr in few hours without affecting the radio-purity of  the LAr.

The tuning is performed by generating simulation samples obtained by choosing optical parameters near their nominal values for comparison to dedicated DarkSide measurements or extracted from published literature. The large number of parameters led to a massive ($\mathcal{O}(10^4)$) collection of  simulation samples. The selected parameters were then determined by simultaneously minimizing the $\chi^2$ of the  four variables listed above.  

\begin{figure}[t]
\centering
\includegraphics[width=.48\textwidth]{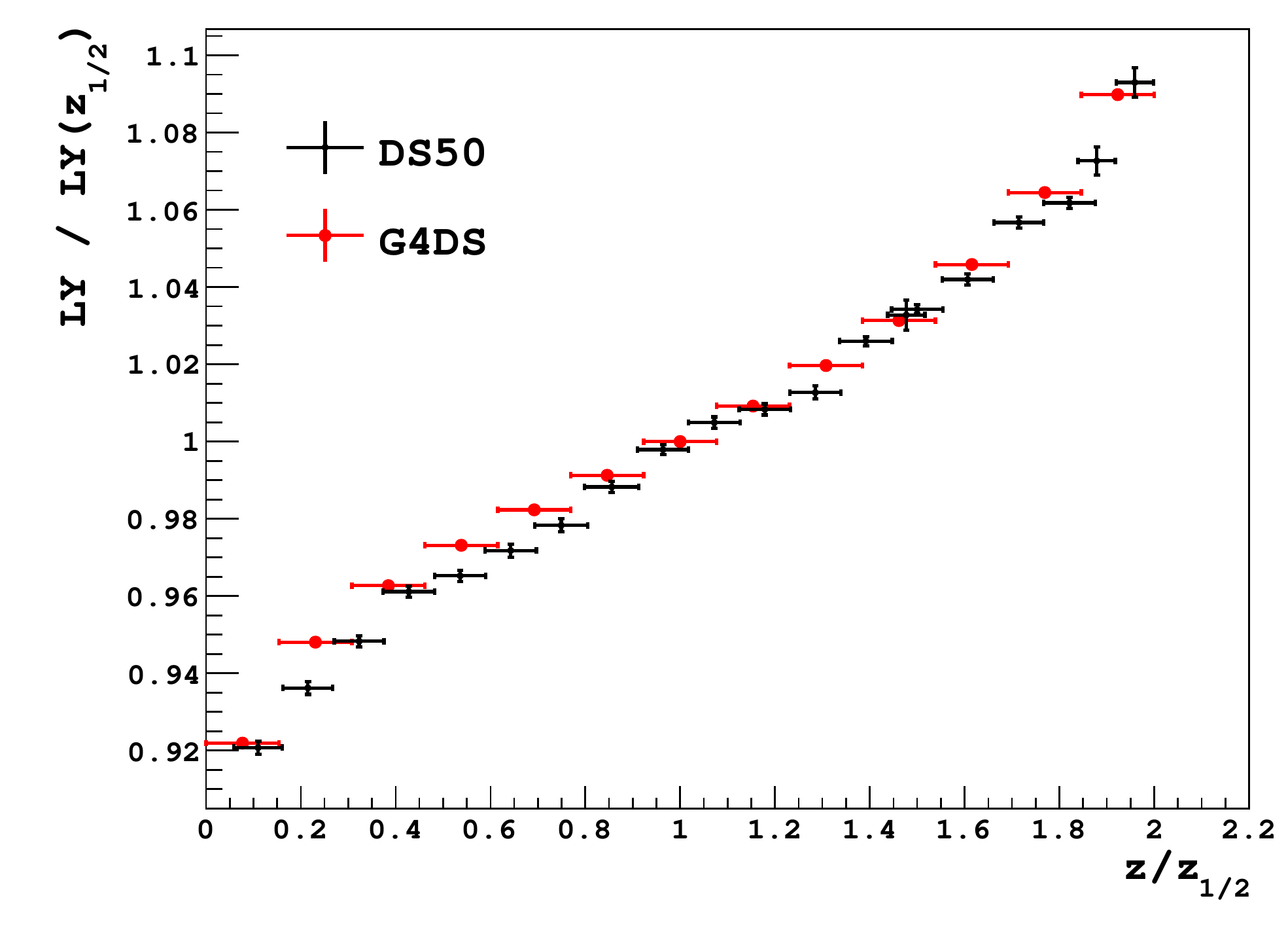}
\includegraphics[width=.48\textwidth]{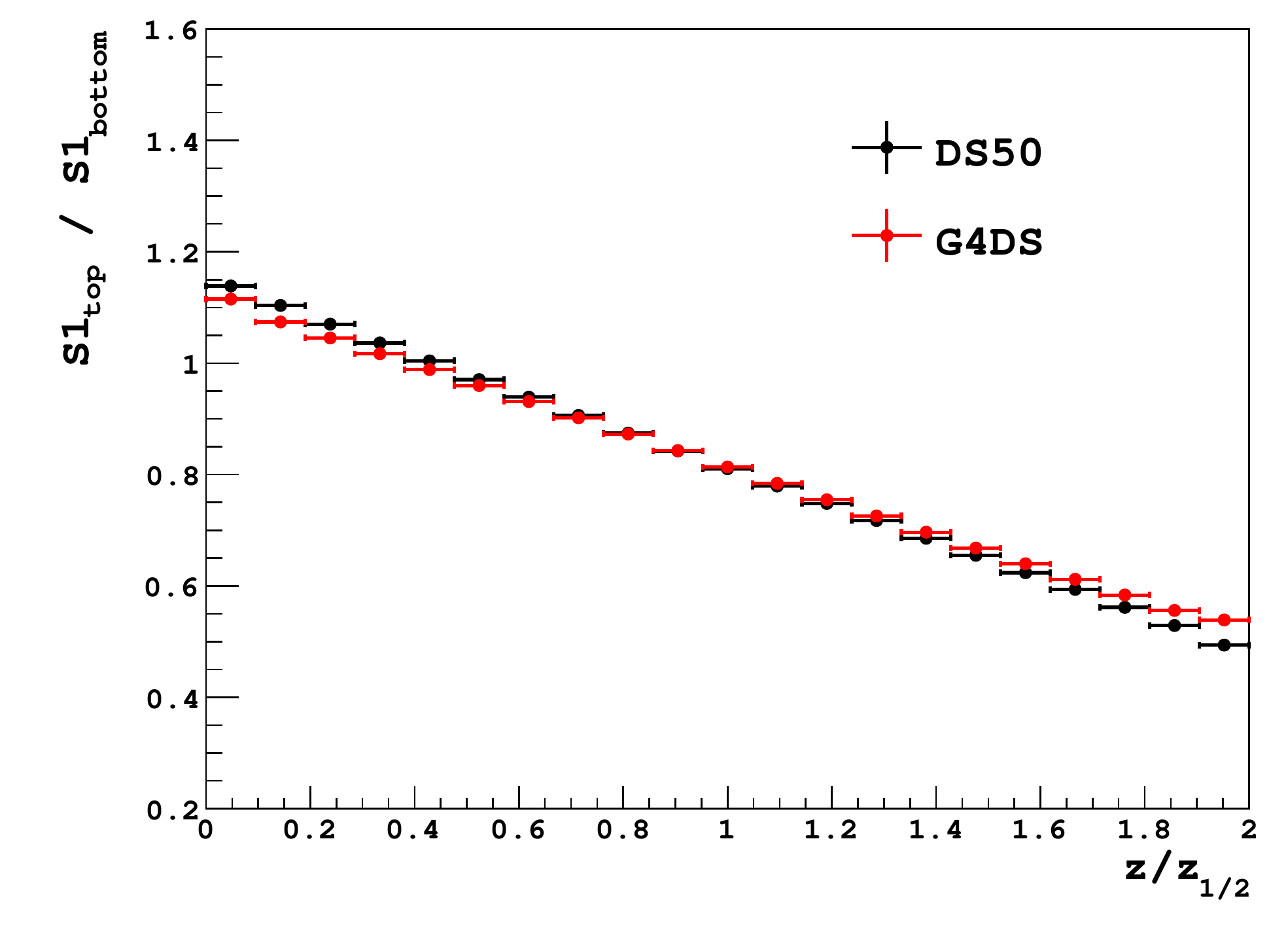}
\mycaption{Relative light collection (left) and  top-bottom ratio in light collection (right)}{as a function of the vertical position of the event. The normalization is relative to the center of the TPC, so that $z/z_{1/2}=0$ corresponds to the top, $z/z_{1/2}=2$ correspond to the bottom.} 
\label{fig:lyvsz}
\end{figure}

With the above geometry the simulation underestimated the top/bottom  ratio observed in the data, and the $\chi^{2}$ minimization was unable to converge. This problem was resolved by introducing  in the simulated geometry a condensed liquid argon layer (100~$\mu$m thick) on the underside of the top fused silica window, as quoted in Table \ref{tab:materials}. This is motivated by considering the possibility that the gas layer was condensing on the window, interfaced with LAr on the other side. An alternative explanation is that the layer represents an empirical correction  for the TPB optical response to visible photons, based on a limited amount of data from lab measurements. The layer of condensed liquid argon provides additional  total internal reflection between liquid  and gaseous argon, and enhances the probability of photon collection in the top PMTs. The presence of this additional layer reduces the disagreement between the data and the MC simulation of the collection efficiency, and the top/bottom ratio as function of z-position to the sub-percent level, as shown in Figure~\ref{fig:lyvsz}. 

\begin{figure}[t]
\centering
\includegraphics[width=0.48\linewidth]{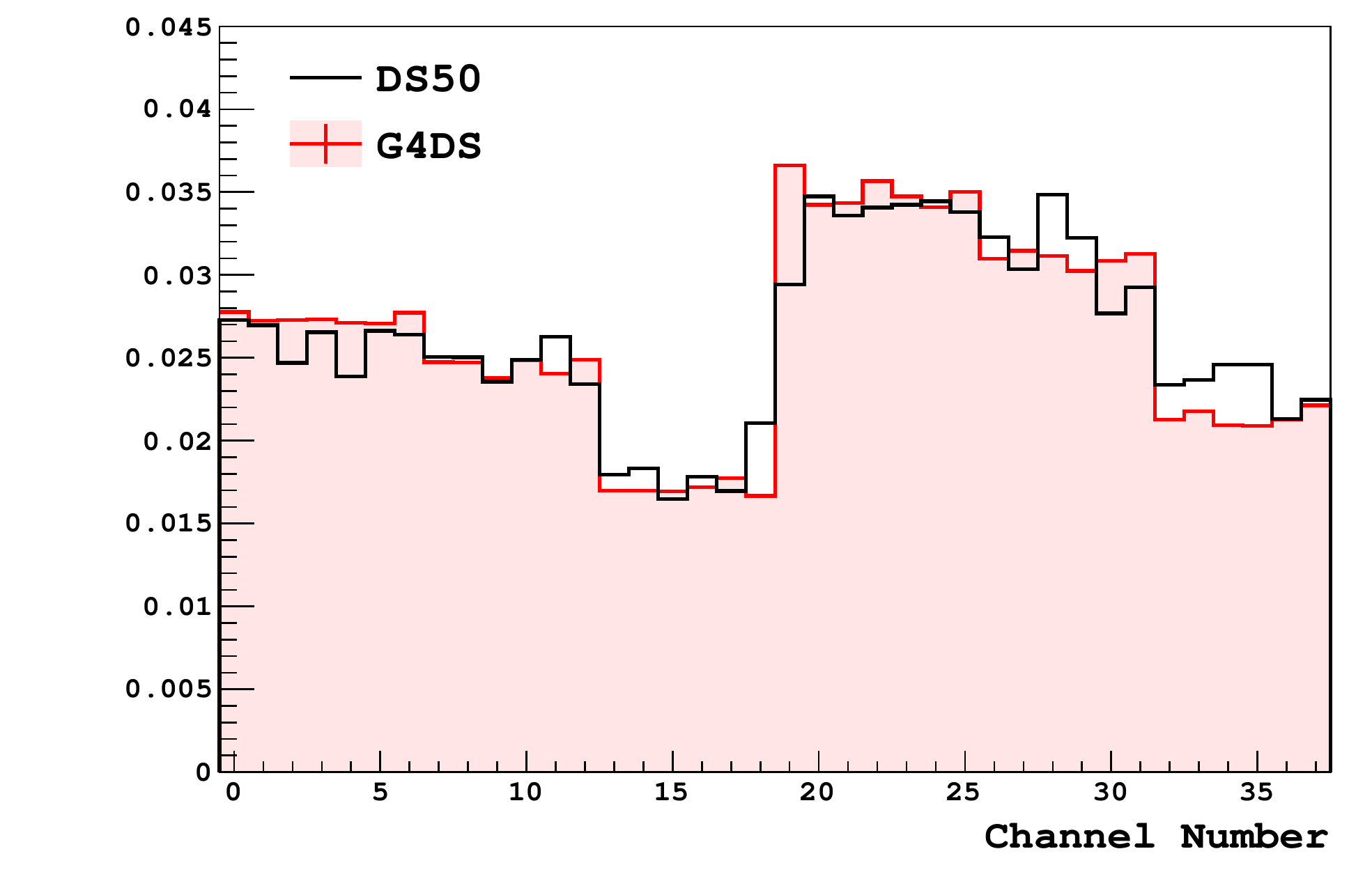}
\includegraphics[width=0.48\linewidth]{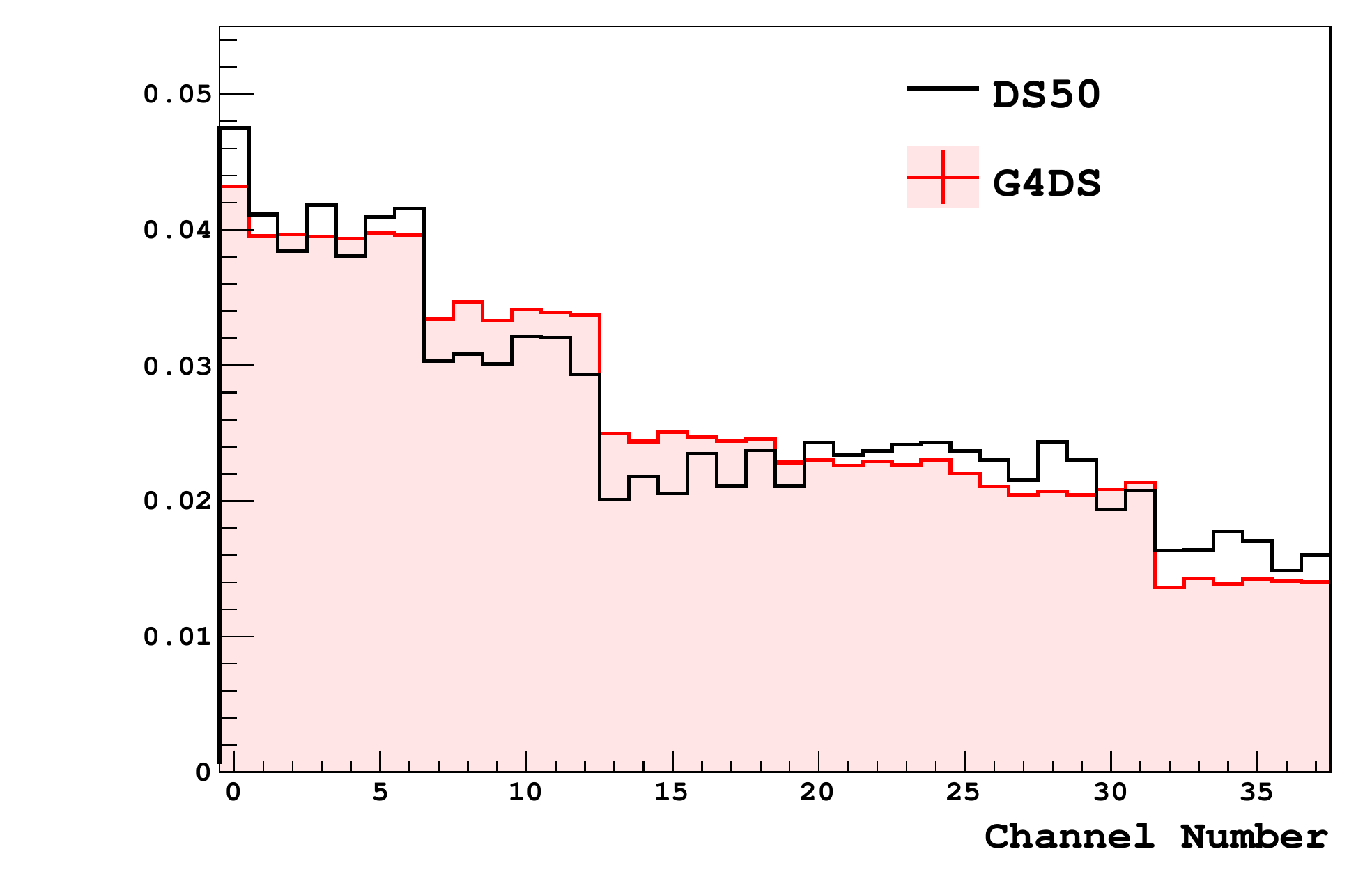} 
\caption{S1 (left) and  S2 (right) channel occupancy comparisons between  data and G4DS. Channels between 0-18 (19-38) correspond to the top (bottom) array of PMTs. } 
\label{fig:light_radial}
\end{figure}

The PMT channel occupancy is the best estimator testing the light collection as a function of the radius in the plane orthogonal to the electric field. The excellent agreement, at a few percent, between S1 data and G4DS for most of the PMTs was verified by the uniformly distributed $^{39}$Ar events (see Figure \ref{fig:light_radial}).   

\begin{figure}[h]
\centering
\includegraphics[width=.6\textwidth]{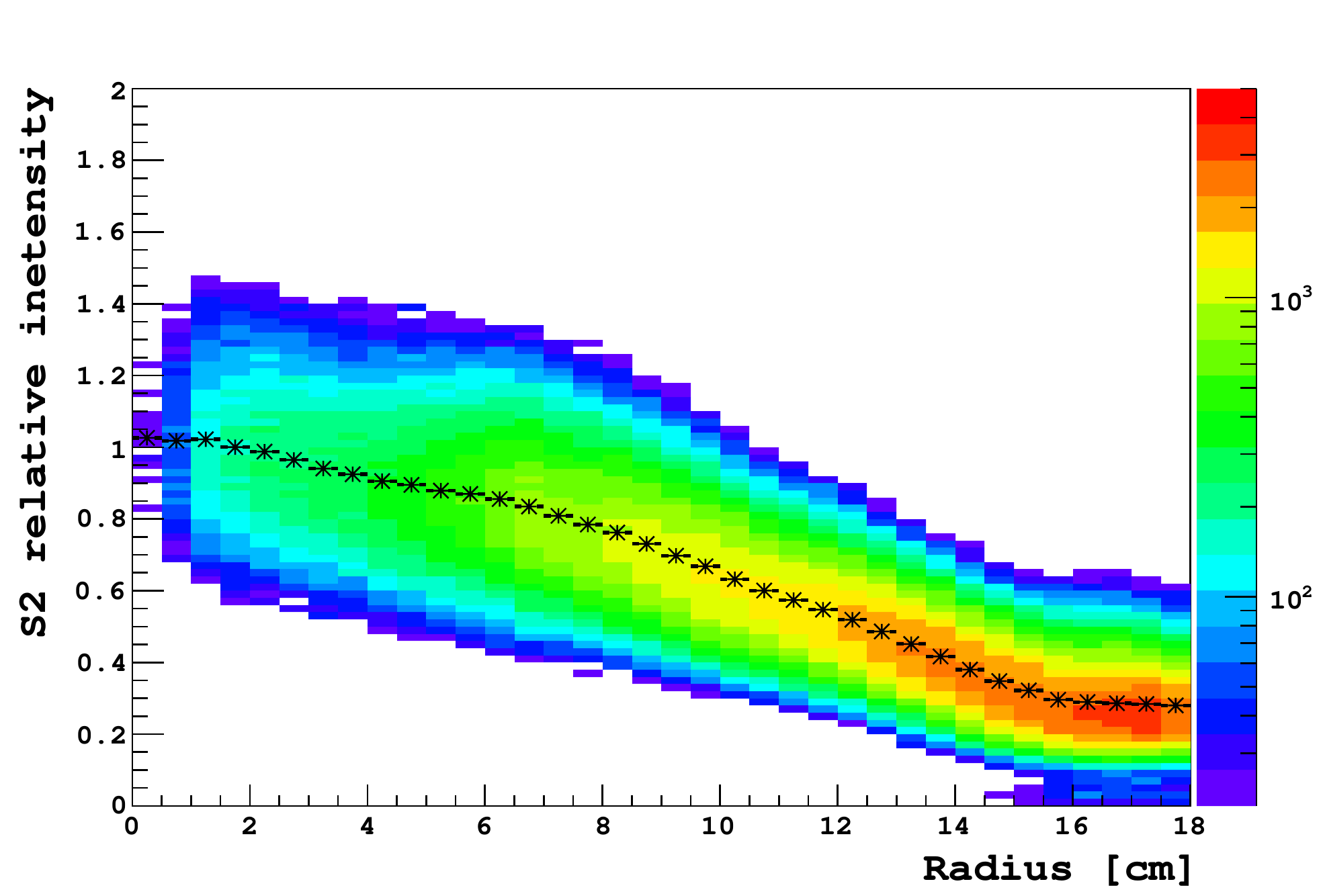}
\caption{Radial dependence of the S2 relative  yield, extracted from $^{83m}$Kr. Black dots correspond to the mean S2 value at fixed radii. }
\label{fig:s2rad}
\end{figure}

S2 modelling has  to also take into account the radial dependence of the S2 scintillation yield. This was measured in DarkSide-50 with $^{83m}$Kr data, which can be  selected by S1 cuts. The xy-position reconstruction, which is based on the S2 pattern on the top array of PMTs, allows extraction of the radial dependence of the S2 yield (Figure~\ref{fig:s2rad}), and reveals a  factor of $\sim$4 difference between the S2 yield at the TPC center compared to the edges. This effect could be produced by a radial non-­uniformity in the electric field in the gas region caused either by a deformation of the top window or a distortion of the grid, or a non-planarity of the grid or the liquid surface. Another potential cause could be non--uniformity of the condensed LAr layer thickness, on the lower surface of the top window.  Once such effects are included in the simulation, the S2 channel occupancy in G4DS agrees with the data, as shown in Figure \ref{fig:light_radial}  (right), within a few percent for most of the PMTs.

Light propagation in LAr also depends on absorption and Rayleigh scattering. LAr absorption has a minimal impact since its characteristic absorption length  (3.8$\pm$1.2~m from the tuning of  G4DS)  is much larger than the  TPC size. On the other hand, DarkSide-50 is very sensitive to Rayleigh scattering, which is modelled in G4DS as function of the wavelength. The method is described in \cite{Seidel2002189, Grace:2015yta}. The  best fit to the data for the scattering length at 128~nm, is 46$\pm$11~cm, which is in agreement with $55\pm5$~cm obtained in reference~\cite{Grace:2015yta} where the Rayleigh scattering length was extrapolated from measurements at higher wavelengths.  However, it is in tension with a value of $66\pm3$~cm, which was directly measured at  128~nm \cite{ISHIDA1997380}, and a value of 90~cm that is obtained from theoretical considerations \cite{Seidel2002189}.


\section{The PARIS model for scintillation and ionization in liquid argon}

\label{sec:paris}
The fine  tuning of the TPC optical response as described in the previous section almost entirely resolves the degeneracies  between the light collection  in DarkSide-50 and the LAr S1 and S2 energy response,  which was expected to be non--linear at non--null fields. G4DS adopts an effective  model  to parameterize the processes inducing the S1 and S2 signals. This model is called PARIS (Precision Argon Response Ionization and Scintillation), which is coded in a single Geant4 process class and relies on the fundamental principles  governing the ionization and scintillation processes of LAr.

A fraction of  the energy deposited  by external radiation in noble liquids is converted  into  $N_i$  electron-ion pairs, and in $N_{ex}$ excited atoms. A residual fraction of energy is dissipated by heating, either by producing   secondary nuclear recoils or inducing  sub-excitation electrons. The combination of these heating processes are grouped under the name of quenching, and is dominant  for nuclear recoils, where the quenching reduces the visible energy by a factor $\sim$3-5 in LAr \cite{Cao:2014gns}, but is negligible for electron recoils. It is assumed in PARIS that  there is no quenching  for electron recoils. Using this assumption, the energy deposited by an   electron recoil, $E_{dep}$, is  divided into excitation and ionization, 

\begin{equation}
E_{dep} = N_i W_i + N_{ex} W_{ex},
\label{eq:work}
\end{equation}

\noindent where  $W_i$ and $W_{ex}$ are the ionization and excitation work functions required  to produce an ion-electron pair and an exciton, respectively.   For simplicity, an effective work function, \textit{W}, is introduced so that

\begin{equation}
E_{dep} = W (N_{i} + N_{ex}).
\label{eq:enevariable}
\end{equation}

\noindent Then defining  $\alpha_k$=$N_{ex}/N_i$, where $k$=\{ER, NR\}  for electron and nuclear recoils respectively, the effective work function can be expressed as

\begin{equation}
W = \frac{\alpha_k W_{ex} + W_{i}}{1 + \alpha_k}.
\label{eq:effective_work}
\end{equation}

\noindent Consequently, the number of electron-ion pairs and excitons can be re-defined as

\begin{equation}
N_i = \frac{E_{dep}}{W} \frac{1}{1 + \alpha_k},
\label{eq:Ni}
\end{equation}

\noindent and

\begin{equation}
N_{ex}= \frac{E_{dep}}{W} \frac{\alpha_k}{1 + \alpha_k}.
\label{eq:Nex}
\end{equation}

The value of $W$ is fixed to 19.5 eV, as suggested by T.~Doke \textit{et al.} \cite{Doke:1988dp}.   The excitation--to--ionization ratio $\alpha_{ER}$ is equal to 0.21 \cite{Doke:1988dp} for electron recoils and $\alpha_{NR}$=1 for nuclear recoils. Since $\alpha_{NR}$ can be fully absorbed by the nuclear recoil quenching function, any deviation from a true $\alpha_{NR}$ value is re-normalized by the quenching factor without loss of generality, as described in Subsection \ref{sec:quenching}.   When ions and electrons recombine, they produce excited atomic states,  which further contribute to the S1 component, and reduce the S2 component.  Each S1 photon can then originate either  directly from the excitation component, or from the recombination of electron-ion pairs. Under the assumption that each excitation and ionization quantum can generate only one photon, S1  can be parametrized as

\begin{equation}
S1 = g_1 (N_{ex} + r(E) \times N_{i}),
\label{eq:S1}
\end{equation}

\noindent where $r(E)$ is the recombination probability, as a function of the kinetic energy $E$,  and $g_1$ is the collection efficiency of photons generated in liquid argon, including the PMT optical coverage and quantum efficiency, and the light absorption in the detector materials.  $g_1$ was estimated equal to 0.157$\pm$0.001, by simulating UV photons uniformly distributed in the TPC.      In a similar way, S2 is proportional to the number of free electrons, which survive recombination and are extracted into the gas pocket
 
\begin{equation}
S2 = g_2  \,  Y_{S2} \times (1 - r(E)) \times N_{i}.
\label{eq:S2}
\end{equation}

\noindent Here, $g_2$ (0.163$\pm$0.001)  accounts for the  detection efficiency of photons generated in the gas pocket and $Y_{S2}$ is the electroluminescence  yield.   In G4DS, all electrons are assumed to  recombine with ions under null  field. With this assumption, S1 in Equation \ref{eq:S1} can be expressed as
\begin{equation}
S1 = g_1 (N_{ex} +  N_{i}) = g_1 \frac{E_{dep}}{W}.
\label{eq:S10}
\end{equation}

\noindent The above assumption is confirmed within 2\% for energies between 40--670 keV by the MicroCLEAN experimental  results \cite{PhysRevC.81.045803}. In the presence of an external electric field, electrons  can escape the ion-pair cloud, reducing the recombination contribution to S1 signal and enhancing the S2 signal.. The probability of escaping the cloud depends on the cloud size, which is defined by the Onsager radius \cite{Onsager:1938zz}, and hence on the stopping power of the recoiling particle.  

Several models have been developed to describe the recombination probability as functions of particle energy and electric field. Among these,  Jaffe's theory   \cite{PhysRev.58.968}  proposes diffusion equations for positive ions  and electrons to explain the columnar recombination of ion-electron pairs around the particle track when subject to an external electric field.  This model was tested  with a $^{207}$Bi source in LAr at different fields by E. Aprile \textit{et al.} \cite{Aprile1987519}. The Thomas-Imel box model  \cite{PhysRevA.36.614} simplifies  Jaffe's theory  by assuming zero diffusion and zero ion mobility. This model matches the data in the regime of "short" tracks, { \it i.e.} when the track is shorter than the mean ionization electron--ion thermalization distance.  This is shown by T. H. Joshi \textit{et al.} \cite{JoshiPhysRevLett.112.171303} for $^{37}$Ar in LAr. The Doke-Birks model \cite{Doke:1988dp}, which is based on Birks' law, empirically parametrizes the recombination effect and was found to be in good agreement with data  in the "long" track regime. M.~Szydagis \textit{et al.} \cite{Szydagis:2011} suggests an approach where both models  are applied, after a  regime selection. This method is embedded in a Geant4-based simulation code called NEST \cite{Szydagis:2011}, and has been successfully applied in several liquid xenon experiments \cite{Aprile:2012nq,Akerib:2013tjd}.

The NEST approach combines  the  Thomas-Imel and Doke-Birks models, by constraining the associated parameters in several experimental data sets.  Such sets are rather abundant  for  liquid xenon but limited to only a few energies for argon (see for instance  \cite{ScalettarPhysRevA.25.2419,Aprile1987519,Sangiorgio201369,Cao:2014gns,1748-0221-12-05-C05010,BONDAR2016119}).  The attempt to implement a NEST--like approach in G4DS suffers from the lack of LAr data, where the difficulty lies in choosing when and how to transition between the two regimes. Thus  PARIS embeds a simplified and effective model in order to reproduce only  DarkSide-50 data. However as shown here, the extremely good agreement found between data and Monte Carlo, suggests that a generalization of the PARIS approach beyond DarkSide-50 is possible.   

The essence of the PARIS approach is an  empirical parametrization of the recombination probability, which depends on the kinetic energy of the ionizing particle. This probability is defined over the entire energy  range for single scatter events, and takes the functional form

\begin{equation}
r(E) = {\rm erf}(E/p_1) ( p_2 \times e^{-E/p_3}+p_4),
\label{eq:reco}
\end{equation} 

\noindent where ${\rm erf}$ is the error function, $E$ is the kinetic energy, and $p_i$, with  $i$ ranges from 1 to 4, are free parameters derived from the fit of  DarkSide-50 data, as   shown in the next subsection. The dependence on the electric field  is not considered since Equation \ref{eq:reco} was tuned only on for  DarkSide-50 data with a 200 V/cm drift field.    

The nuclear recoil energy  is defined with respect to the electron recoil scale after the addition of a quenching factor  independent of the electric field. In the Subsection \ref{sec:quenching}, data and Monte Carlo comparisons are shown for  different  nuclear recoil quenching models. The quenching model in PARIS acts by scaling $E_{dep}$, and hence reducing   the number of quanta,  in Equations \ref{eq:Ni} and \ref{eq:Nex}.  The tension between the published results of nuclear recoil quenching in LAr, especially at low energies \cite{Creus2015}, suggests implementing the two  most used models, the Lindhard \cite{Lindhard} and Mei  \cite{Mei200812} models. A full description of quenching  modelling in PARIS and the data vs. Monte Carlo comparison are  shown in Subsection \ref{sec:quenching}.


Statistical fluctuations of S1 and S2 are introduced in PARIS only at the photon emission level. In particular, it is assumed that each exciton and each recombined pair can induce exactly one photon. The number of emitted photons contributing to the S1 pulse is modelled with a Poisson distribution, with mean number equal to $N_{ex} + r(E) N_i$. Fluctuations in S2 are modelled by randomly generating photons with Poisson statistics distributed around $Y_{S2}$ for each extracted electron (see Subsection \ref{sec:S2}). The optical properties of  G4DS, described   in the previous section and  independently tuned on the energy response of DarkSide-50 to $^{39}$Ar and $^{83m}$Kr, naturally determine  the light collection efficiencies $g_1$ and  $g_2$ in Equations  \ref{eq:Ni} and \ref{eq:Nex}.  The accurate determination of the detector response as a function of the energy and position of the event, is essential in order to test the robustness of the G4DS approach, as demonstrated in the following subsection.



S1 and S2 photons are emitted with Gaussian distributed wavelengths using $\mu$=128~nm and $\sigma$=2.6~nm.  Singlet and triplet    de-excitation times are fixed to  6~ns and 1.6~$\mu$s, respectively. The  population probability of  the  singlet or triplet state is  determined by a singlet-to-triplet ratio, depending on  the ionizing particle energy.  This is tuned using calibration data (see   Subsection \ref{sec:f90}).   Electrons surviving recombination  are drifted in the MC to the level of the liquid/gaseous argon interface with a velocity of 0.93 mm/$\mu$s. Drifted electrons undergo diffusion, with transverse and longitudinal components of 4.8~cm$^2$/s \cite{Amoruso200468} and 18~cm$^2$/s \cite{689434}, respectively. An electron drift lifetime of 15.8 ms, as measured by DarkSide-50 with $^{83m}$Kr calibration data, is introduced to take into account  the effects of impurities in LAr.  Each electron  is tracked along the drift path in the liquid, and the electron  mobility in gaseous argon is fixed at 475 cm$^2$/s/volt \cite{PhysRev.166.871}.  At the liquid gas interface, the extraction and the  electroluminescence fields are  2.8~kV/cm and 4.2~kV/cm, respectively, and S2 photons are generated along the electron drift path in the gas pocket. The time difference between  PMT detection of S2 photons and particle interaction is the composition of  the electron drift times in  LAr and in gas pocket, the de-excitation of the  excited argon dimer, and the photon time of flight.  

The following subsections discuss in detail the S1 and S2 energy response, the S1 time profile which is crucial for pulse shape discrimination between electron and nuclear recoils, and the quenching of the nuclear recoil energy.



\subsection{The S1 response to electron recoils}\label{sec:S1}

The S1 response to electron recoils is derived  from Equations \ref{eq:Ni}, \ref{eq:Nex}, and \ref{eq:S1}

\begin{equation}
S1 = g_1 \frac{E_{dep}}{W}\frac{r(E) + \alpha_{ER}}{1+\alpha_{ER}}.
\label{eq:finalS1}
\end{equation}

\noindent Ideally, the only undefined quantity in this equation is the recombination probability, $r(E)$, which requires tuning using the data.   However, inaccuracies in  TPC optical tuning can induce   a small degeneracy  between the collection efficiency $g_1$  and the effective work function $W$ of Equation \ref{eq:effective_work}. A scaling factor, whose deviation from one is comparable with  the inaccuracy in collection efficiency,  is expected. Tuning of  the recombination probability relies on the simultaneous fit of the  energy spectra induced by the decays of  $^{37}$Ar, $^{83m}$Kr, and $^{39}$Ar (from the atmospheric argon run), which are present or artificially produced and distributed in the active mass. Cosmogenic $^{37}$Ar decays via  electron capture, emitting an X-ray of 2.62 keV with 8.4\% Branching Ratio (BR). The $^{37}$Ar sample used in this analysis was identified  in early runs of the underground argon campaign, since it  has a relatively short half life ($\tau_{1/2}$=35.5~d).     


$^{39}$Ar is a unique first-forbidden $\beta$-decay, and the deviation of the spectral shape from an allowed $\beta$ transition has to be  taken into account using the  shape factor.  
\begin{equation}
F (T) =  \left( (T + m_e)^2  - m_e^2 \right ) +  ( Q - T )^2.
\end{equation}

\noindent Here T is the kinetic energy, and Q the Q-value, which is known with 1\% accuracy.  Uncertainties on the screening correction   \cite{Hime2011Proc} and on the shape factor   prevent us for calibrating the S1 response at low energies ($<10$~keV) with  $^{39}$Ar. The low energy calibration thus relies on $^{37}$Ar (2.62 keV) and on $^{83m}$Kr (41.5 keV). 

Calibration data for the three sources are selected by applying a fiducialization along the drift field (40 < drift time < 334.5 $\upmu$s) to minimize the contamination from the $^{40}$K, $^{238}$U, and $^{232}$Th from the PMT glass and stem, as done in the  DarkSide-50 analyses for  WIMP searches \cite{Agnes:2014bvk, Agnes:2015ftt}. Background subtractions are applied to the   $^{37}$Ar   and  $^{83m}$Kr data,  using data selected from  periods where the source was not present or negligible. 

For the  $^{39}$Ar sample, the uncertainties on the theoretical spectral shape  impose  a low energy threshold in the fit of this source conservatively set    at 500 pe ($\sim$70~keV). We used the underground argon spectrum  as background for the  atmospheric $^{39}$Ar sample. The only difference between the two, with the exception of the   depletion in the $^{39}$Ar contamination, is an additional $^{85}$Kr contamination in the underground argon sample not observed in atmospheric argon \cite{Agnes:2015ftt}.  This was estimated to be present with an activity of $\sim$2~mBq/kg,  about 0.2\% of  the $^{39}$Ar rate in atmospheric argon and hence neglected in this work. 



\begin{figure}[t]
\begin{center}
\includegraphics[width=.48\textwidth]{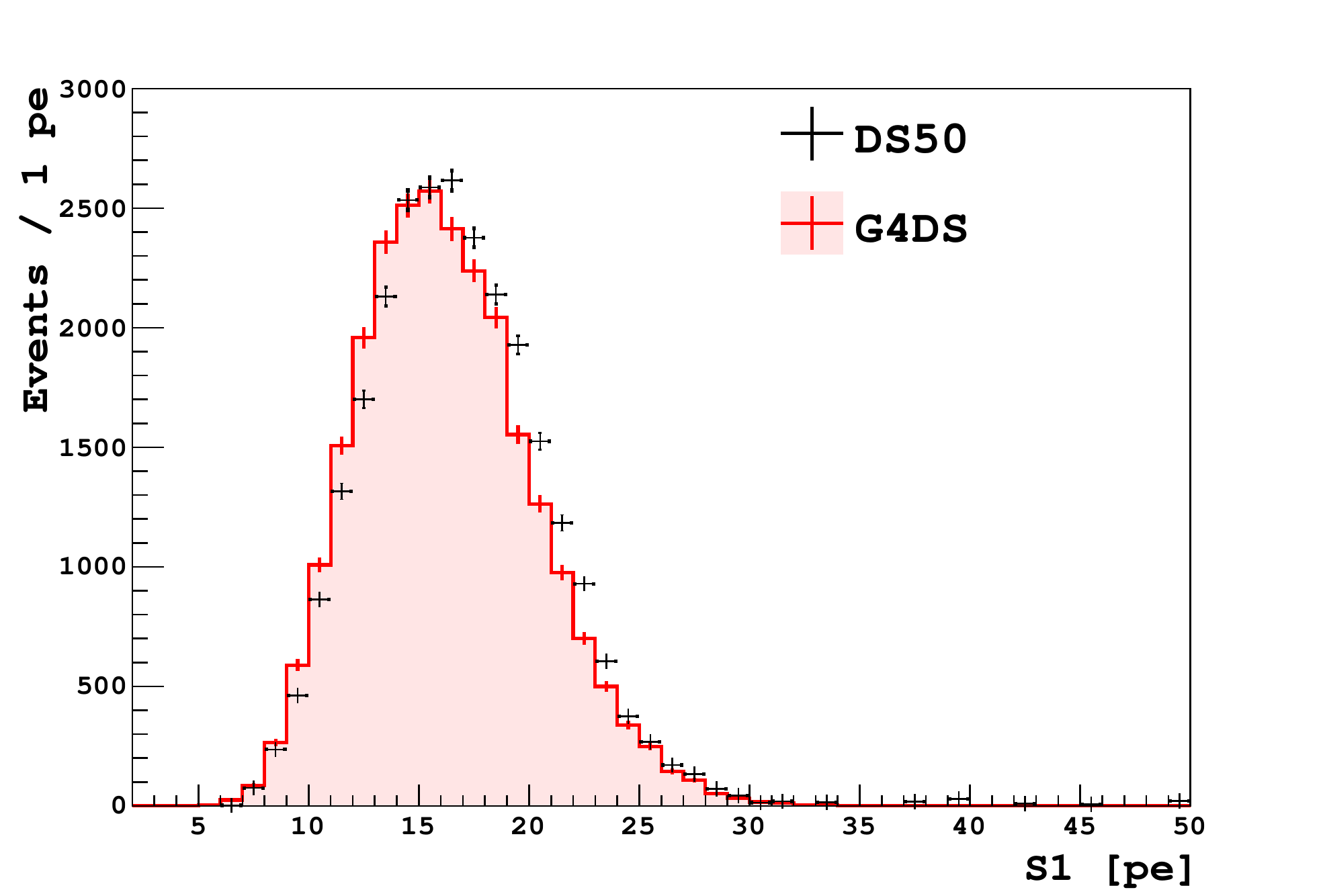}
\includegraphics[width=.48\textwidth]{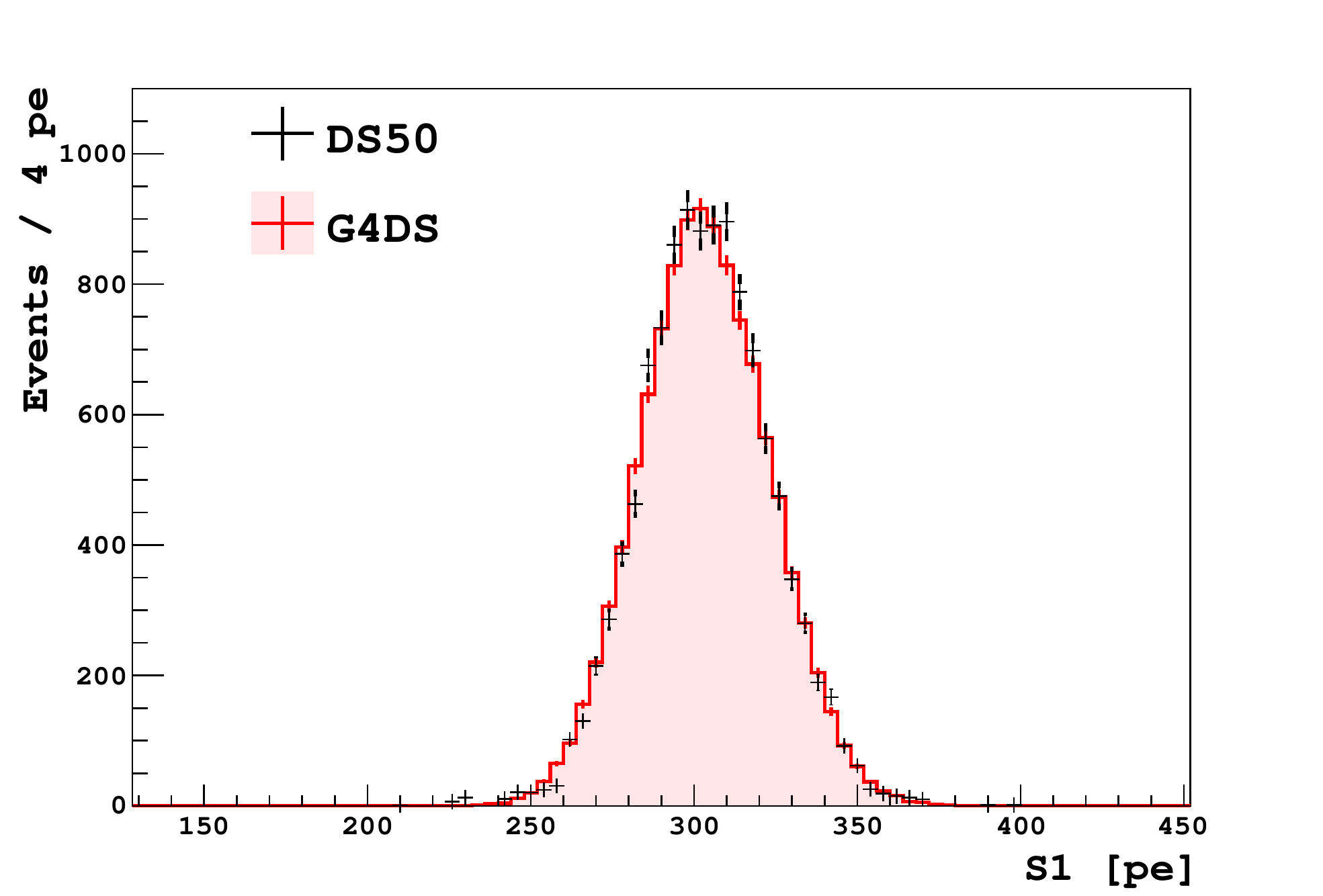}
\caption{Comparison between data (black) and Monte Carlo (red) for $^{37}$Ar (left) and $^{83m}$Kr (right) S1 spectra.} 
\label{fig:s1kr}
\end{center}
\end{figure}

\begin{figure}[h]
\begin{center}
\includegraphics[width=0.65\textwidth]{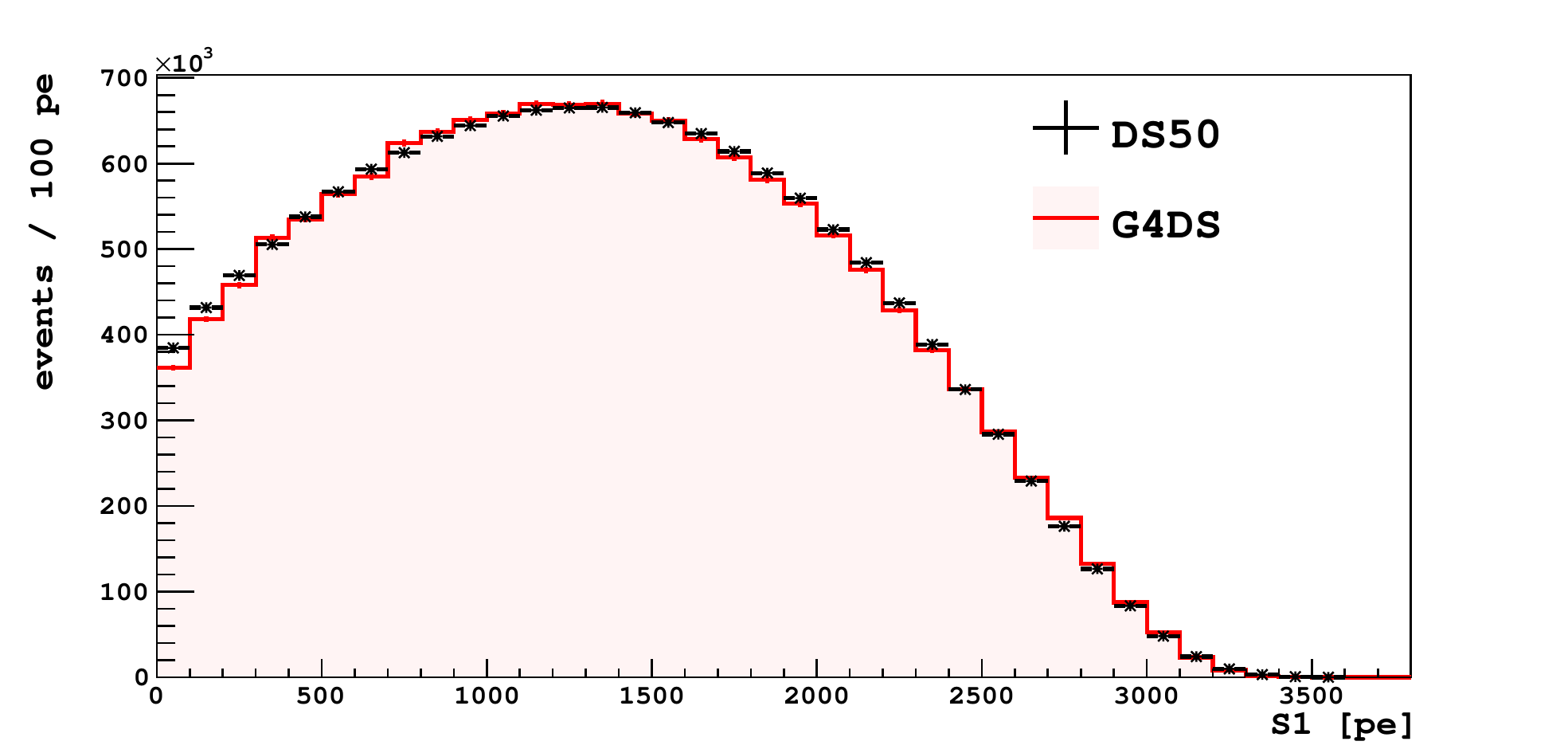}
\caption{Comparison  between data (black) and Monte Carlo (red) for  the $^{39}$Ar S1 spectrum. The energy region below 500 pe was not used to fit the energy scale,  due to the uncertainties on the theoretical spectral shape.} 
\label{fig:s1ar}
\end{center}
\end{figure}

The fit gives in a  scaling factor of 1.01, which is applied to S1 to take into account inaccuracies in the photon collection  efficiency.  The comparisons between data and Monte Carlo are shown in  Figures \ref{fig:s1kr} and \ref{fig:s1ar}.  No corrections are applied to the S1 resolution: the excellent agreement in the S1 resolution between data and Monte Carlo confirms the robustness of the PARIS approach and suggests that the photon emission process is dominated by Poisson statistics.   

\begin{figure}[h]
\begin{center}
\includegraphics[width=.49\textwidth]{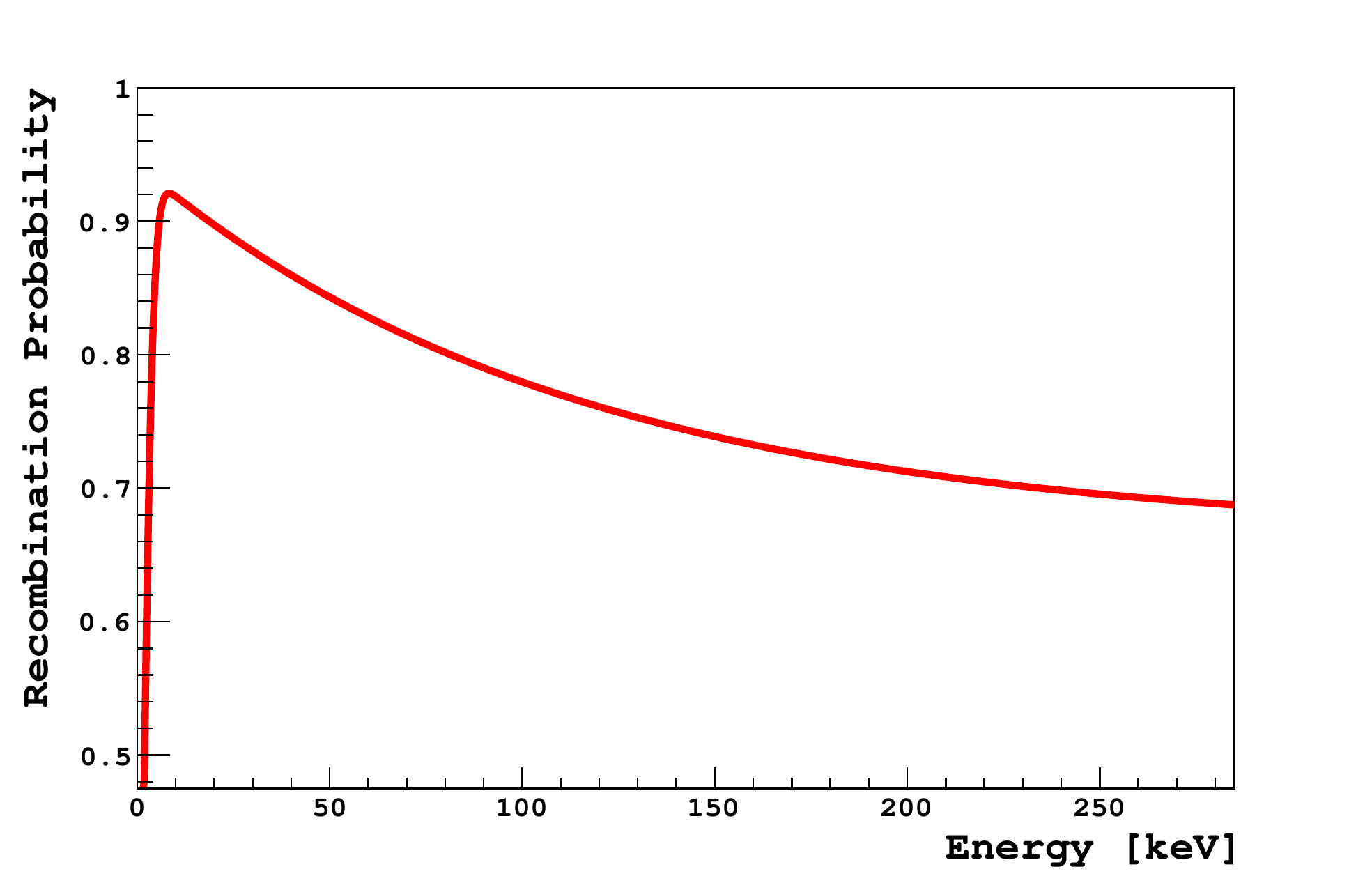}
\includegraphics[width=.49\textwidth]{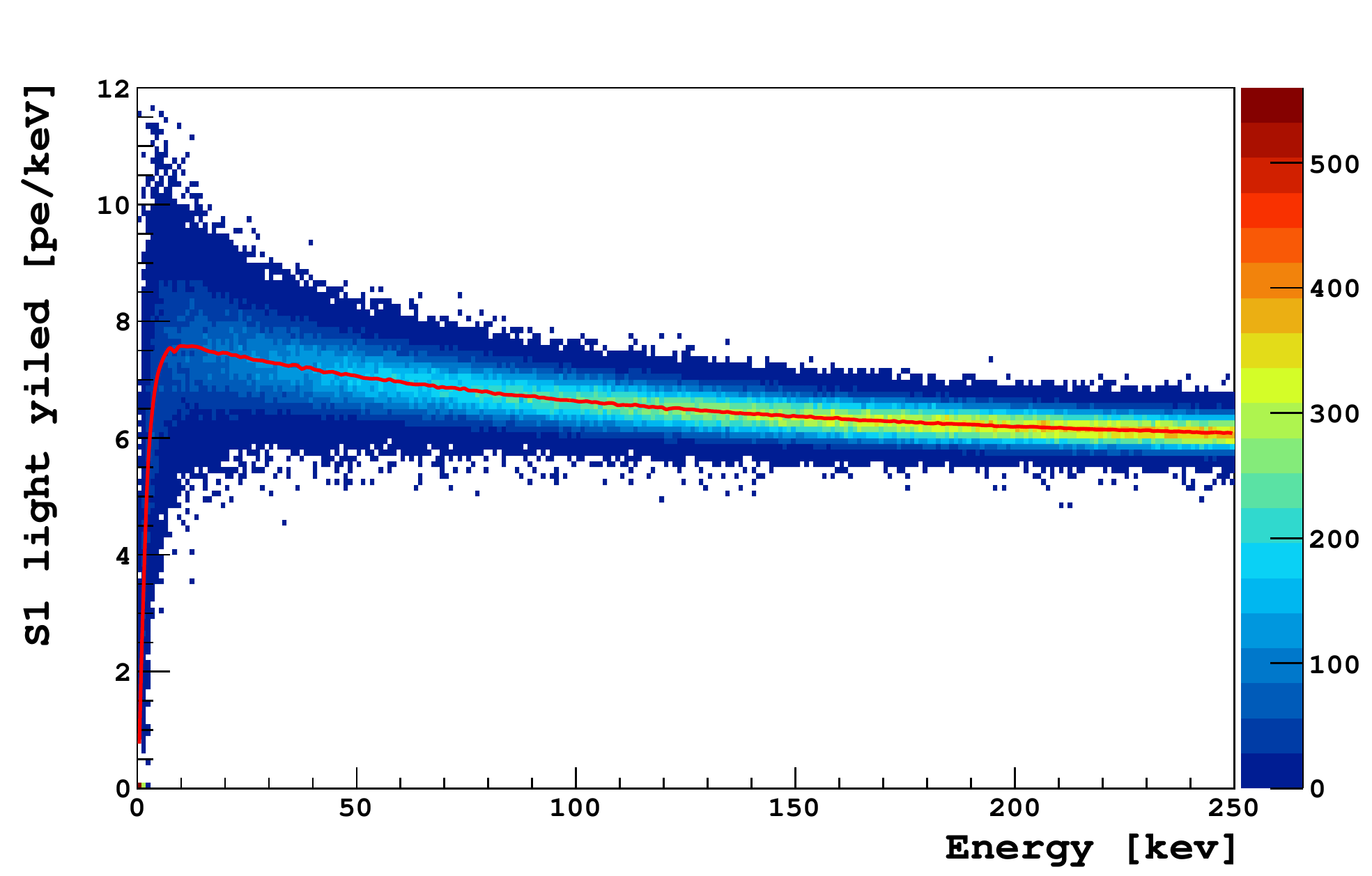}
\caption{Recombination probability as a function of the particle energy, for single scattered events,  extracted from DarkSide-50  data (left) and the G4DS light yield at 200 V/cm (right), defined as S1/E. The red curve in the right panel represents the mean light yield as function of the electron recoil energy. } 
\label{fig:recop}
\end{center}
\end{figure}

The parameters of Equation \ref{eq:reco} obtained from the fit   are  $p_1$ = 3.77,  $p_2$ = 0.277, $p_3$ = 113, and $p_4$ = 0.665 and the corresponding recombination probability is shown  in  Figure \ref{fig:recop} (left) as function of the kinetic energy. At an electric field of 200V/cm, the light yield, defined as defined as S1/E (see Equation \ref{eq:finalS1}), has a maximum at $\sim$15 keV with a value of  $\sim$7.6 pe/keV, and reaches a plateau at high energies equal to $\sim$6 pe/keV (see Figure \ref{fig:recop} right).

In addition to the use of diffuse sources in the TPC, two external  sources, $^{57}$Co  and $^{133}$Ba, were positioned in the liquid scintillator veto close to the cryostat. These were used to cross check the tuning of G4DS.  The dominant gamma line emissions from $^{57}$Co and $^{133}$Ba are at 122 keV and 356 keV,   respectively. These gammas have to cross a double--steel cryostat layer, the bath of liquid argon surrounding the TPC, and the TPC materials (teflon and rings of copper), before reaching the active volume. The non-uniformity in the position distribution of the induced electron recoils impacts the light collection efficiency, which is strongly dependent on the event position. Accurate  calibrations of the optical properties and of the S1 response to uniformly distributed sources naturally overcome this issue.

\begin{figure}[h]
\begin{center}
\includegraphics[width=.48\textwidth]{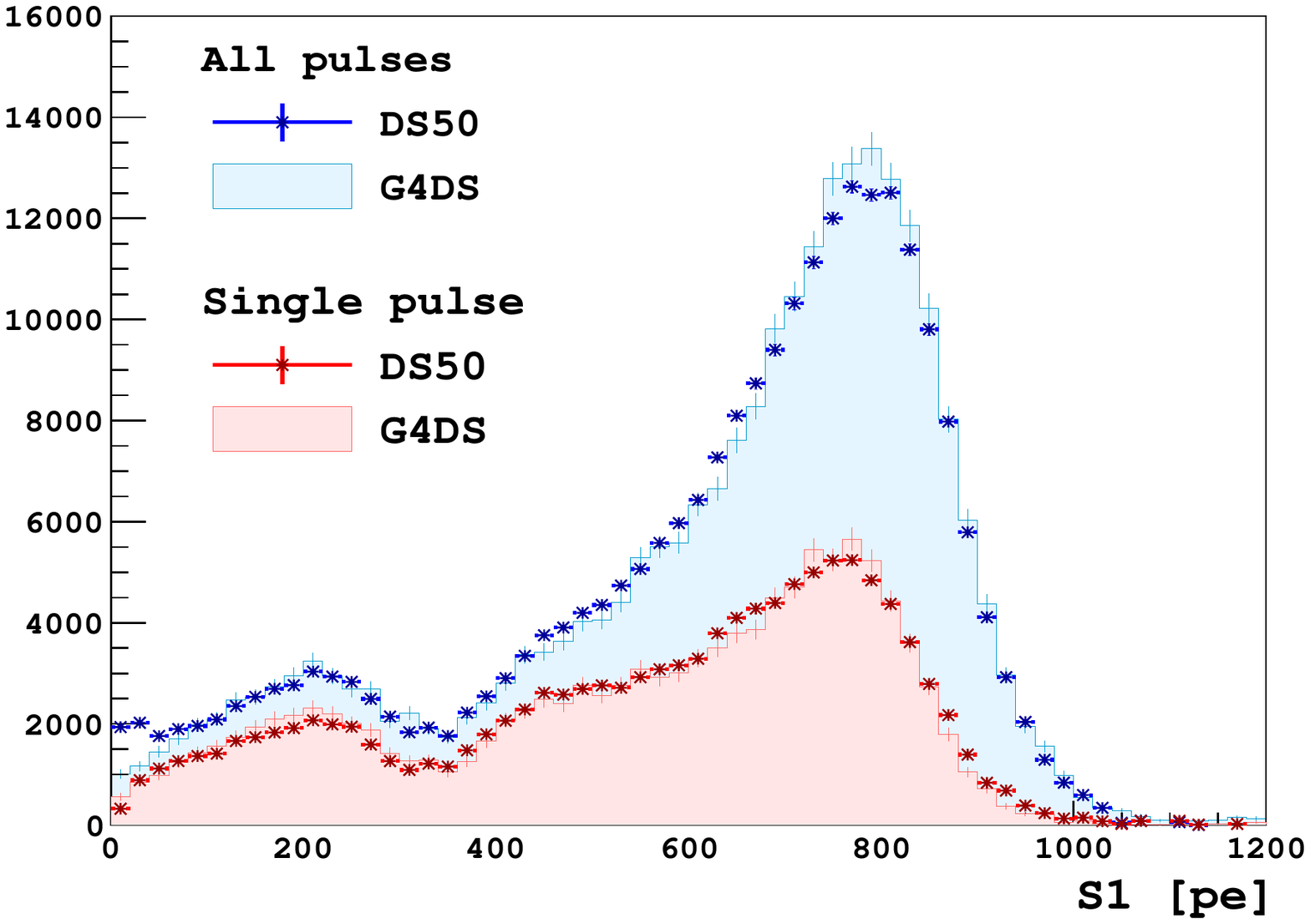}
\includegraphics[width=.48\textwidth]{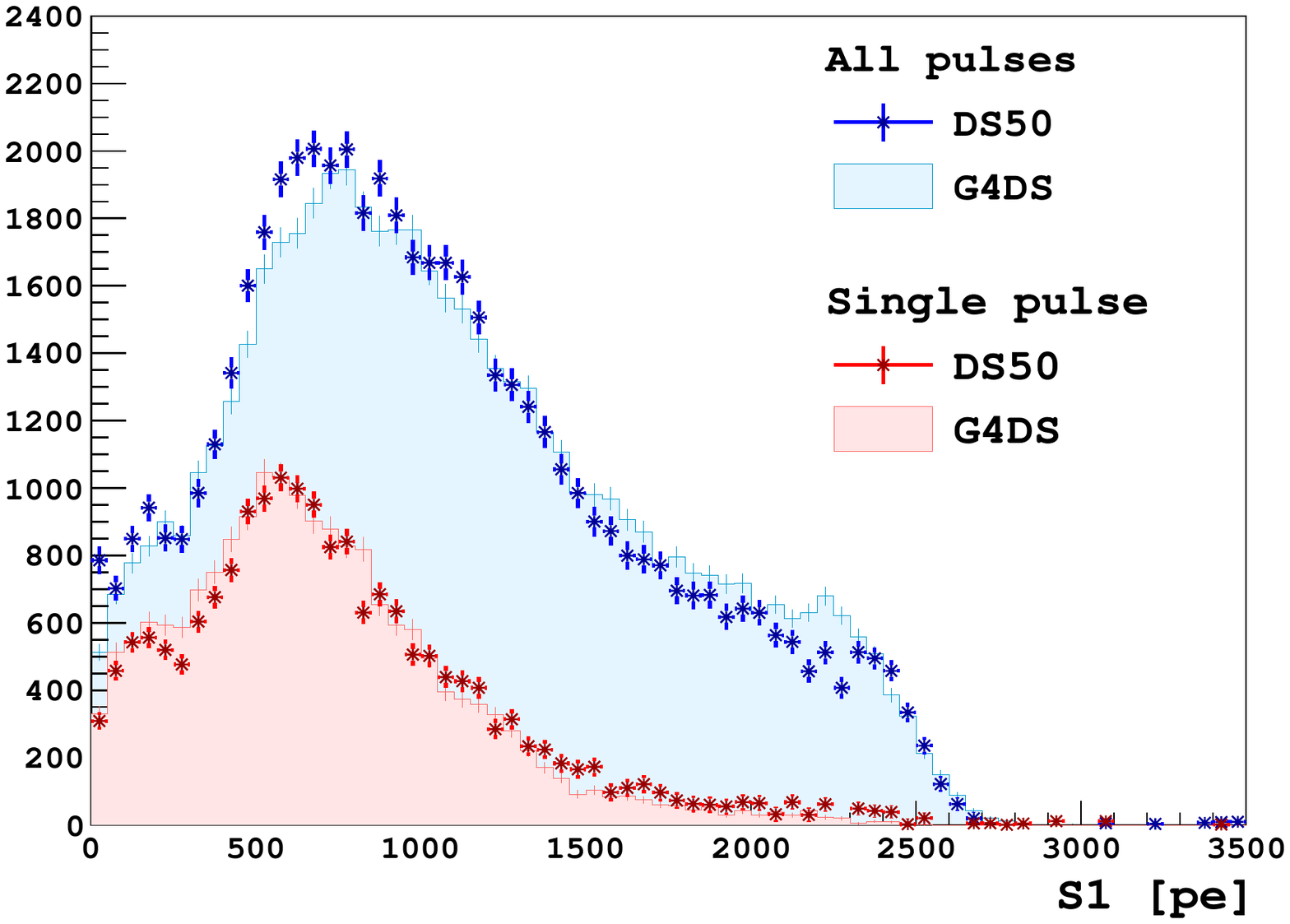}
\caption{Single and multiple scatter S1 spectra of $^{57}$Co (left) and $^{133}$Ba (right) calibration sources in DarkSide-50 data, after background subtraction, and G4DS.} 
\label{fig:calg4ds}
\end{center}
\end{figure}

Figure \ref{fig:calg4ds} shows the excellent  agreement with the energy response of  single scatter events. Since the experiment is designed to look for WIMPs which are also single scatter, this is a critical element for agreement. A small difference in the S1 response was expected for multiply scattered events, when the drift path of ionization electrons cross ion-electron pair clouds from a different interaction. Though this effect is not included in G4DS, the good agreement between Monte Carlo and the $^{57}$Co and $^{133}$Ba calibration sources for multiple scatter events, as shown in Figure \ref{fig:calg4ds},  demonstrates that   interference between ion-electron pair clouds   is negligible for a drift field of 200 V/cm.  

\noindent A further test on the validity of the PARIS model is provided by the ratio of $g_1$/$W$
\begin{equation}
\frac{g_1}{W} =  \frac{S1}{E_{vis}},
\label{eq:invertedS1}
\end{equation}

\noindent where 
\begin{equation}
E_{vis} =   E_{dep} \frac{r(E) + \alpha_{ER}}{1+\alpha_{ER}}. 
\label{eq:Evis}
\end{equation}

 
\noindent The ratio, $g_1$/$W$,  being independent of the field,  allows to extrapolate the recombination effect at  null field. In the limit case of  $r(E)$=1,  where
 
\begin{equation}
\frac{g_1}{W} =  \frac{S1}{E_{dep}},
\label{eq:invertedS1_2}
\end{equation}
 
 \begin{figure}[h]
\begin{center}
\includegraphics[width=.48\textwidth]{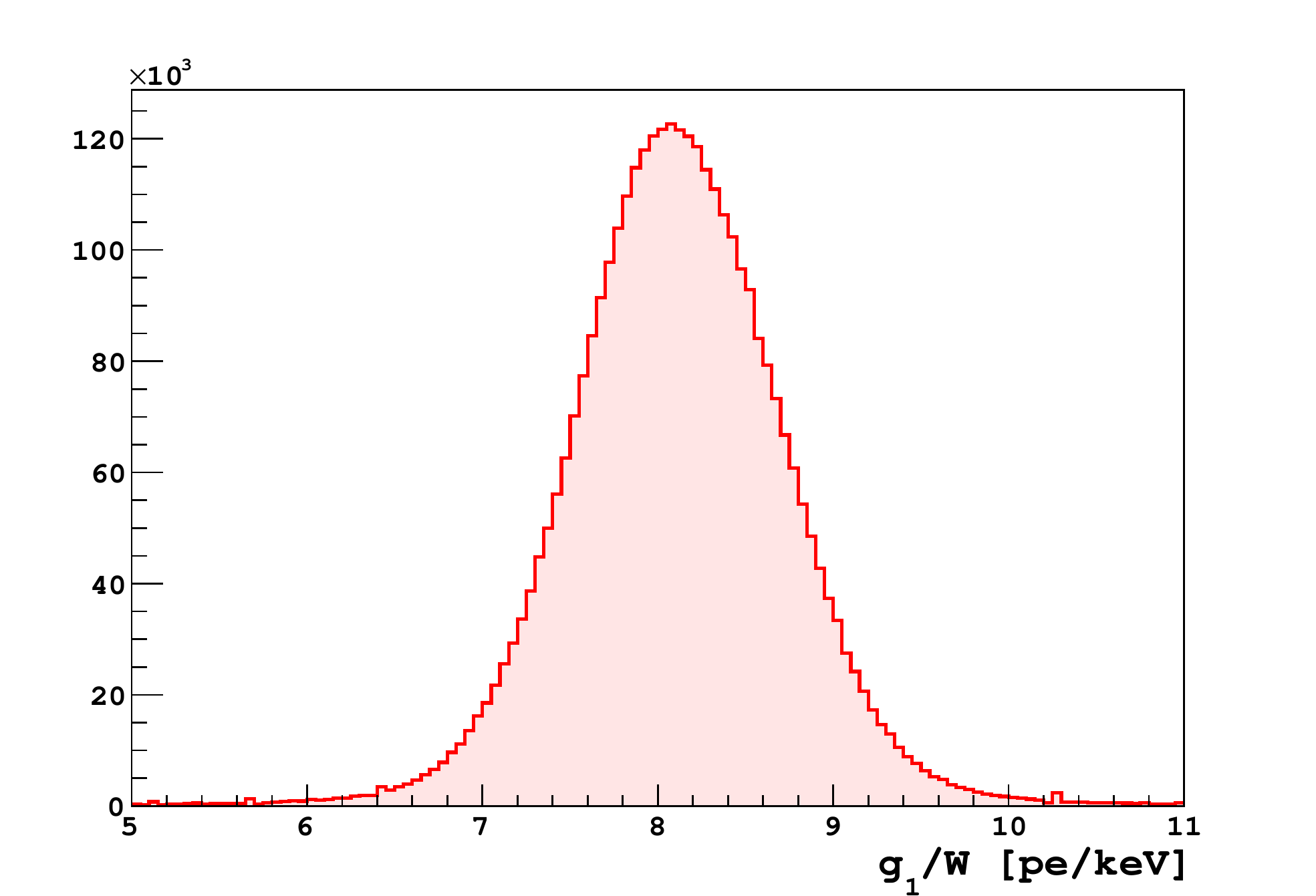}
\includegraphics[width=.48\textwidth]{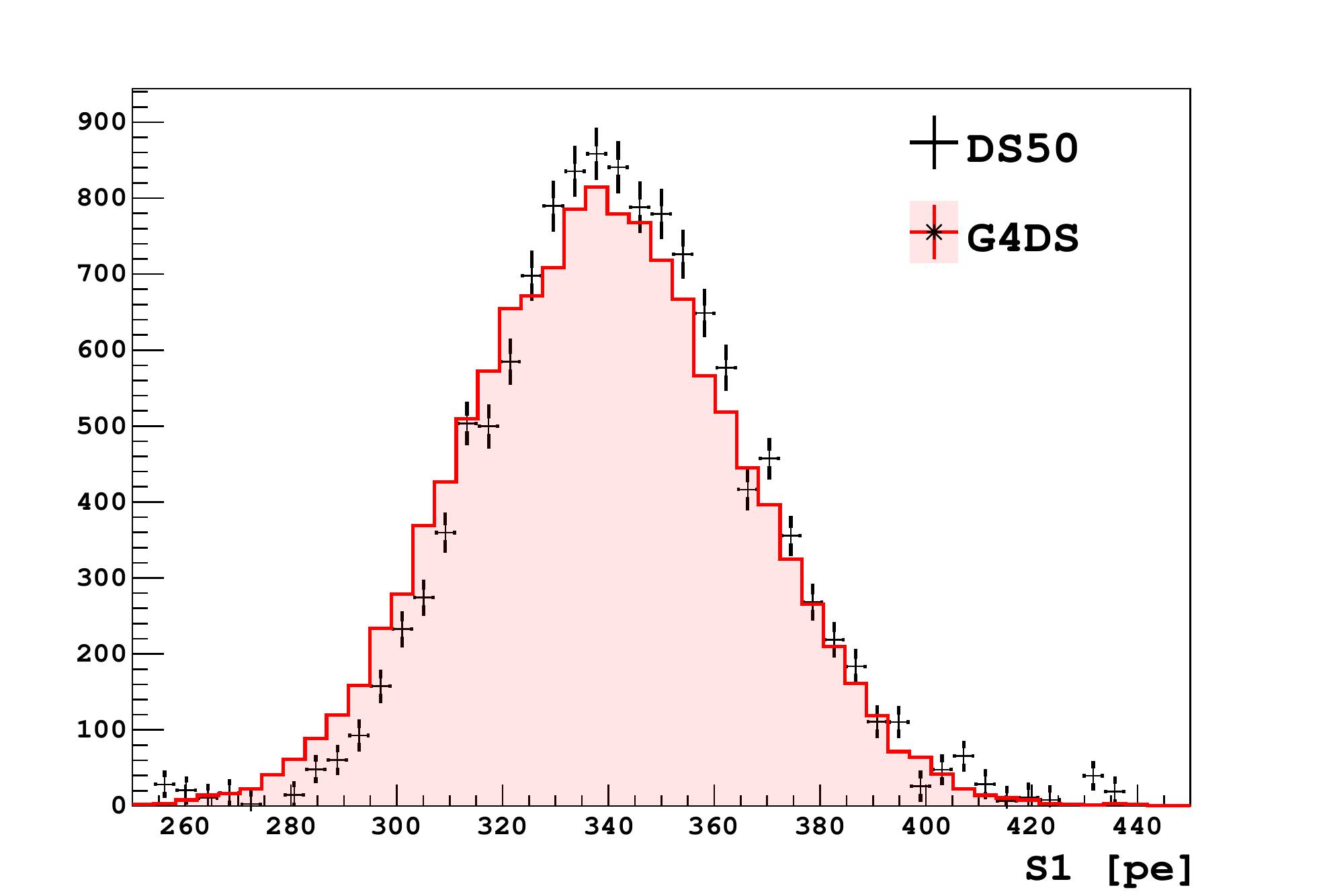}
\caption{Left:  $g_1$/$W$  from the simulation of $^{39}$Ar events at  200 V/cm. The mean value, 8.11~pe/keV, is in good agreement with the light yield, 8.1$\pm$0.2 pe/keV,  measured at null field. Right: comparison of the S1 response to  $^{83m}$Kr between  data and Monte Carlo at zero field. } 
\label{fig:g1w}
\end{center}
\end{figure}

 
\noindent   G4DS  predicts the light yield to be 8.11 pe/keV.  Figure  \ref{fig:g1w} shows the $S1/E_{vis}$ distribution   of  simulated   $^{39}$Ar decays  at 200 V/cm (left) and the data/Monte Carlo  comparison of $^{83m}$Kr  at null field (right).  The measured value  of 8.1$\pm$0.2 pe/keV in DarkSide-50   with a $^{83m}$Kr  source, after the correction for the z-dependence \cite{Agnes:2015ftt}, is in excellent agreement with the G4DS prediction, suggesting that most of the electron-ion pairs recombine at null field.  This extrapolation is, however, in conflict with the biexcitonic model \cite{PhysRevB.46.11463} that predicts  S1 quenching at null field due to  electrons escaping the  spatial distribution of ion-electron pairs along a particle track.


\subsection{The S1 pulse shape profile}\label{sec:f90}

As already mentioned, the pulse shape profile  is dominated by the characteristic de-excitation times of the singlet ($\tau_s$ $\sim$6~ns) and triplet ($\tau_t$ $\sim$1.6~$\mu$s) states of the argon dimers and by the probability to populate the various states.  Electron recoils induce a  slower pulse, since they mostly populate the triplet state, contrary to nuclear recoils which have a larger probability to populate the singlet state. Other sub-dominant effects affecting the pulse shape profile, like the TPB absorption-emission mechanism,  the photon propagation, the electronics noise, and the PMT jitter are intrinsically provided by  G4DS and by the electronics simulation.

The DarkSide-50 estimator for the S1 pulse shape discrimination, $f90$, is the ratio between the integral of the light pulse in the first 90 ns  with respect to the total integration of up to  7~$\mu$s.

\begin{figure}[h]
\begin{center}
\includegraphics[width=.49\textwidth]{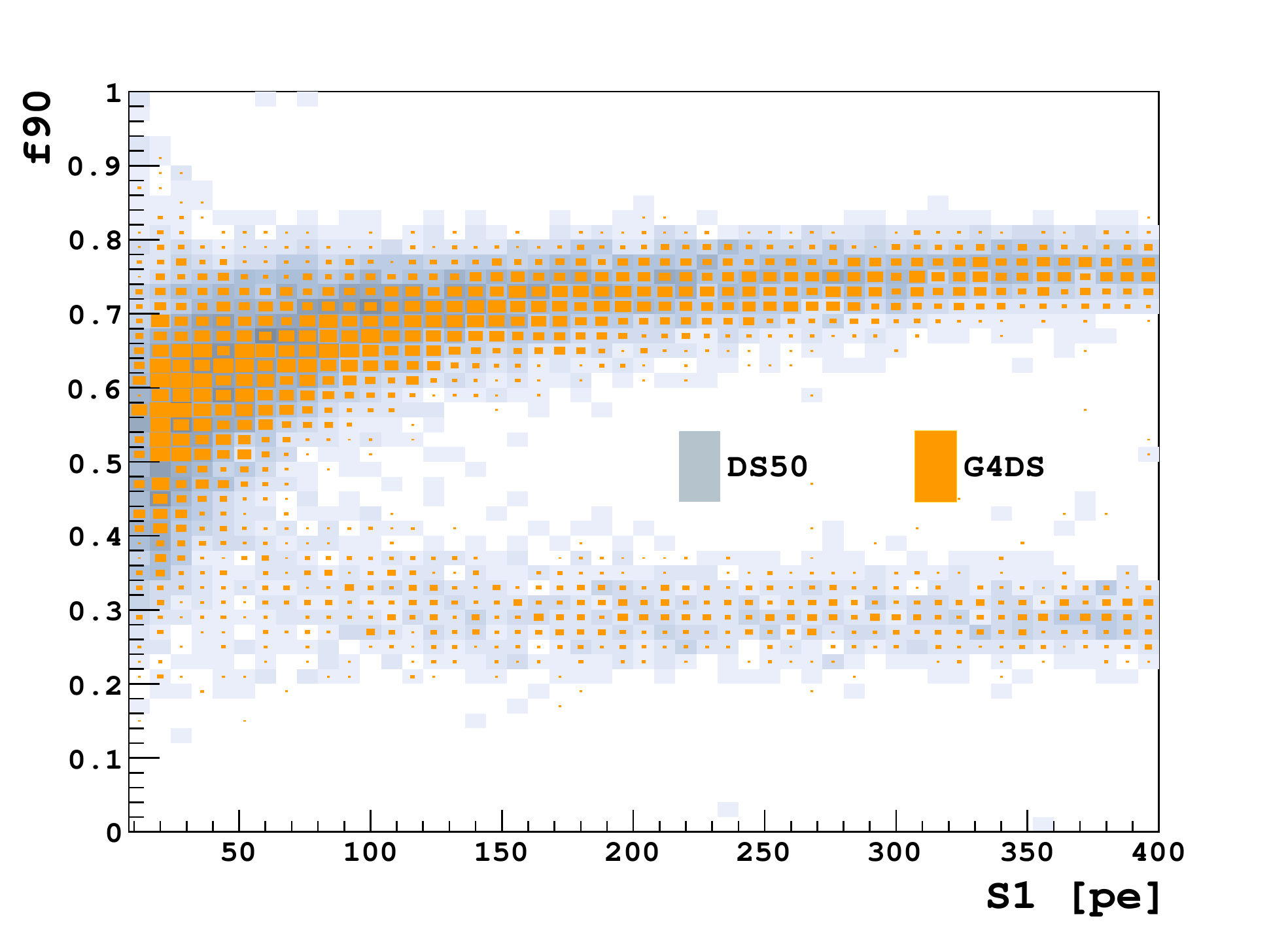}
\includegraphics[width=.49\textwidth]{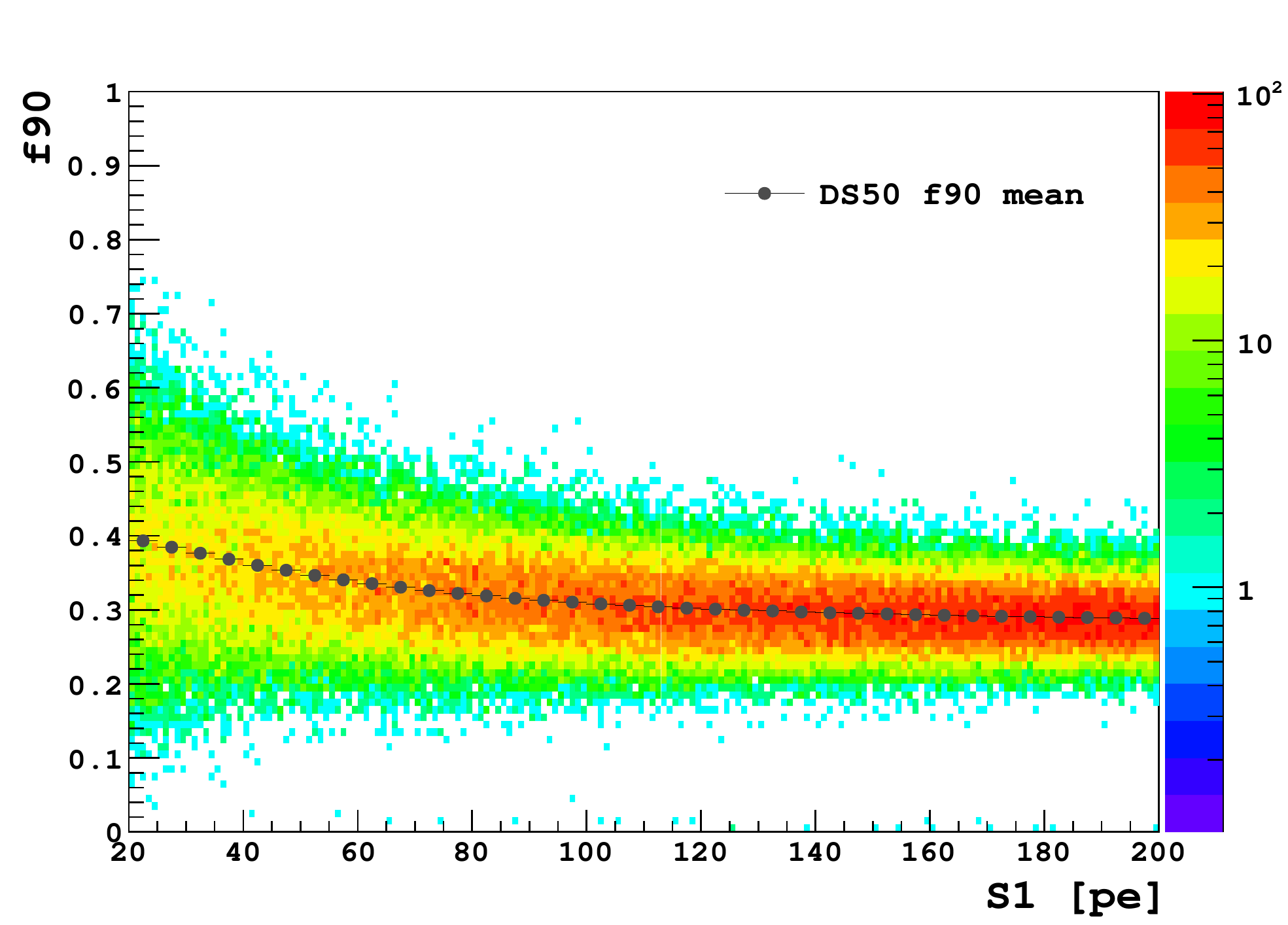}
\caption{Left: $f90$ distribution as a function of S1 for  the AmBe neutron source. Right:  $f90$ vs S1 for simulated  $^{39}$Ar  electron recoils   compared with mean $f90$ extracted from the data. } 
\label{fig:f90mc}
\end{center}
\end{figure}

This ratio is simulated in G4DS by assigning each excited dimer to the singlet (triplet)  state using  a  binomial distribution with  probability $p_s$ (1 -$p_s$).  The probability  at 200 V/cm as  function of the recoil energy, $p_s$,  is derived   for  electron recoils using $^{39}$Ar $\beta$ decays and AmBe for nuclear recoils induced by scattered neutrons. The mean value of the $f90$ distribution was obtained by  fitting  this distributions for each 1--pe bin of S1 with the Hinkely model \cite{PhysRevC.78.035801} and converted in the  $p_s$/(1-$p_s$) ratio ($f_{S/T}$)  by fixing $\tau_s$ and $\tau_t$  to 6~ns and 1.6~$\mu$s, respectively. The comparisons between  data  and Monte Carlo  in Figure \ref{fig:f90mc} illustrate good agreement for $f90$ mean values for both nuclear and electron recoils.

\begin{figure}[h]
\begin{center}
\includegraphics[width=.49\textwidth]{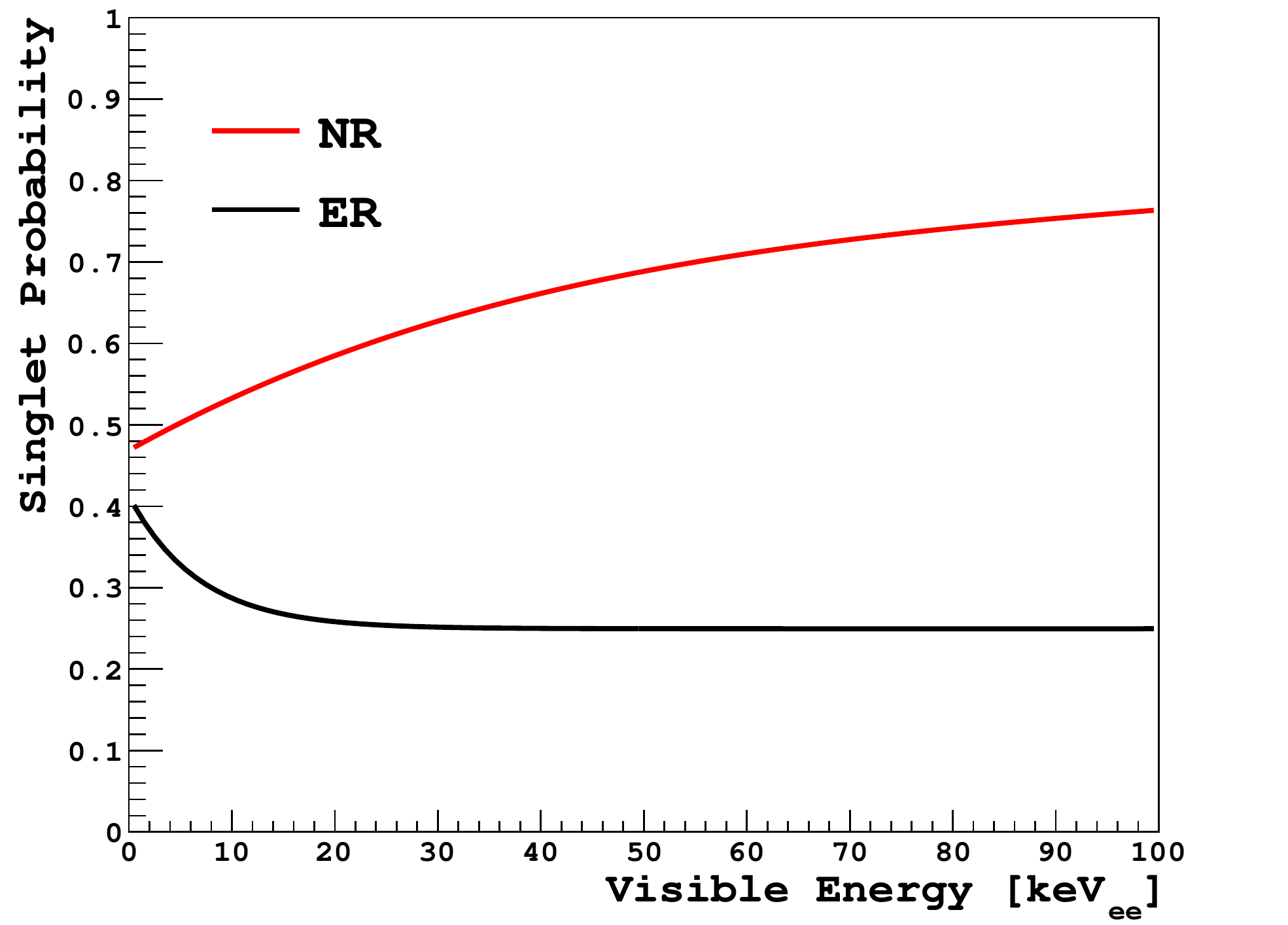}
\caption{Singlet probability, $p_s$, for electron and nuclear recoils as a function of the visible energy at 200 V/cm drift field as coded in G4DS.} 
\label{fig:truef90}
\end{center}
\end{figure}

\begin{figure}[h]
\begin{center}
\includegraphics[width=.49\textwidth]{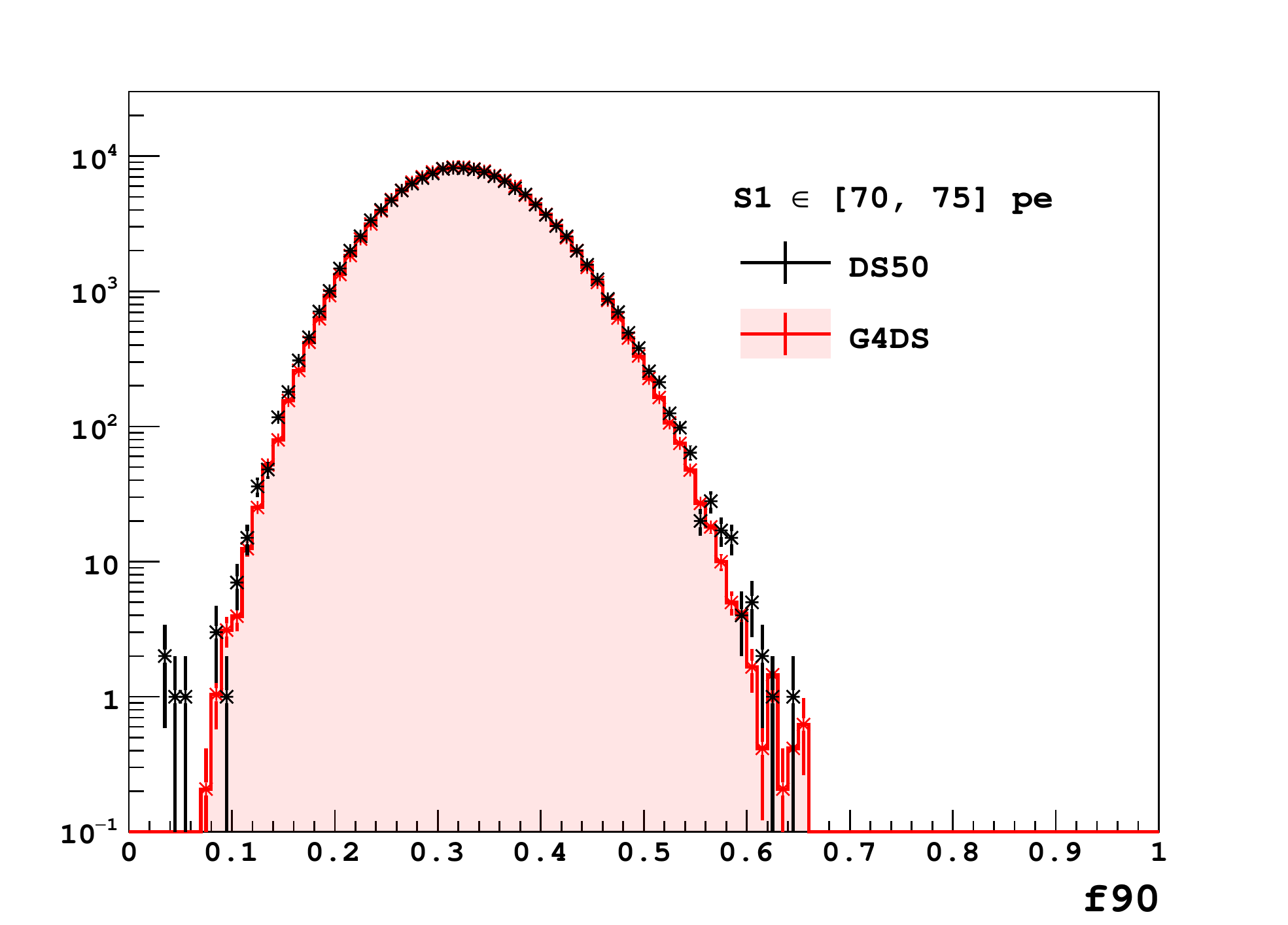}
\includegraphics[width=.49\textwidth]{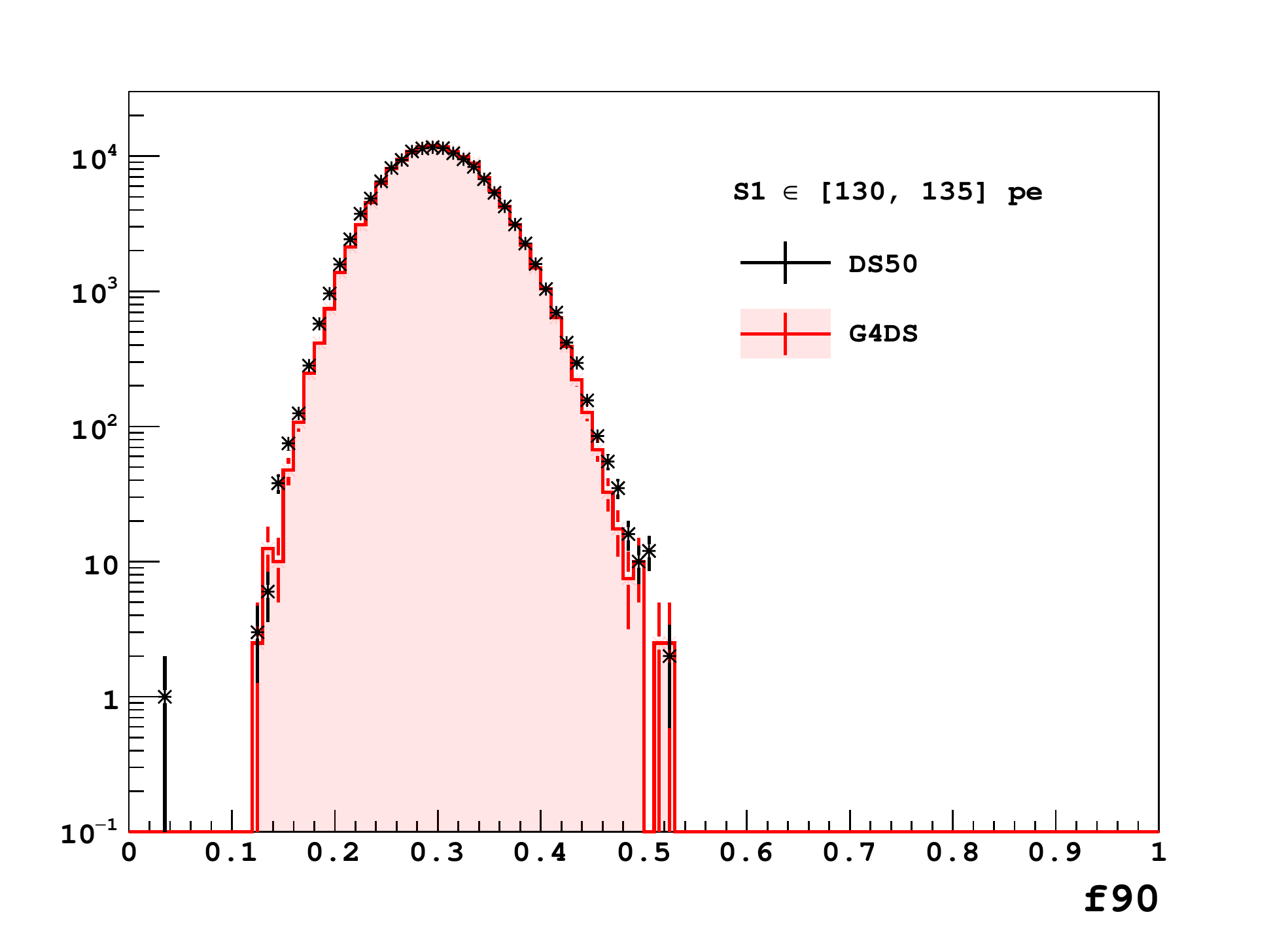}
\caption{DarkSide-50 $f90$ distributions from the atmospheric argon campaign compared to  G4DS simulations for  S1 in [70, 75] pe (left) and [130, 135] pe (right) intervals. The selected sample includes  single scatter events with a cut in the drift time within  40~$\mu$s and  330~$\mu$s.} 
\label{fig:f90g4ds}
\end{center}
\end{figure}

Figure \ref{fig:truef90} shows the extracted singlet probabilities for electron and nuclear recoils at 200 V/cm. The two $p_s(E)$ curves for electron and nuclear recoils are coded in the PARIS model.   The relevant result of this work  demonstrates the capability of G4DS to very accurately predict the electron recoil $f90$ statistical fluctuations, a key ingredient to define the WIMP acceptance band in the DarkSide-50  experiment. The $f90$ distributions are shown in Figure \ref{fig:f90g4ds} for two different S1 ranges, [70, 75] pe and [130, 135] pe, for the high statistics atmospheric argon data. In these two particular cases, the sample statistics count more than 10$^5$ events and correspond to $\sim$20  years of DarkSide-50 operations with underground argon.  


\subsection{The quenching of nuclear recoils}\label{sec:quenching}

The quenching factor in liquid argon (often called $L_{eff}$ or scintillation efficiency\footnote{$L_{eff}$ is usually defined with respect to  the electron recoil light yield at null field at 122 keV, the gamma line emitted by $^{57}$Co. The other definition for the quenching factor, the so-called  Lindhard factor, is defined as the ratio between S1 and the nominal nuclear recoil energy at null field. In G4DS we confirm the linearity at null field for the electron recoil S1 energy scale, as observed by MicroCLEAN, at 2\%. The two definitions provide the same information within this uncertainty.}) was measured by SCENE \cite{PhysRevD.88.092006}, MicroCLEAN \cite{PhysRevC.85.065811}, WARP \cite{Brunetti2005265}, and W.~Creus \textit{et al.} \cite{creus:in2p3-01163988}. The four measurements agree within 2~$\sigma$ with a value of approximately 0.28 at energies above 50  keV. At low energies, however,  there is a tension between the SCENE  data set, where  $L_{eff}$ decreases down to $\sim$0.22 at 11 keV, and MicroCLEAN and W.~Creus \textit{et al.}, who observed an increase to 0.35 (15 keV) and 0.315 (16 keV), respectively.

In order to account for the nuclear recoil quenching,  G4DS provides the option to switch between the   Lindhard theory \cite{Lindhard} and the D. M. Mei \textit{et al.} \cite{Mei200812} model, which modifies the Lindhard expression  as follows

\begin{equation}
L_{eff}^{M}= L_{eff}^{L} \times \frac{1}{1 + k_B \frac{dE}{dx}},
\label{eq:mei} 
\end{equation} 

\noindent where $L_{eff}^{L}$  is the Lindhard factor \cite{Lindhard}  and $k_B$ = 7.4$\times$10$^{-4}$ MeV$^{-1}$ g cm$^{-2}$.











\begin{figure}[h]
\begin{center}
\includegraphics[width=.66\textwidth]{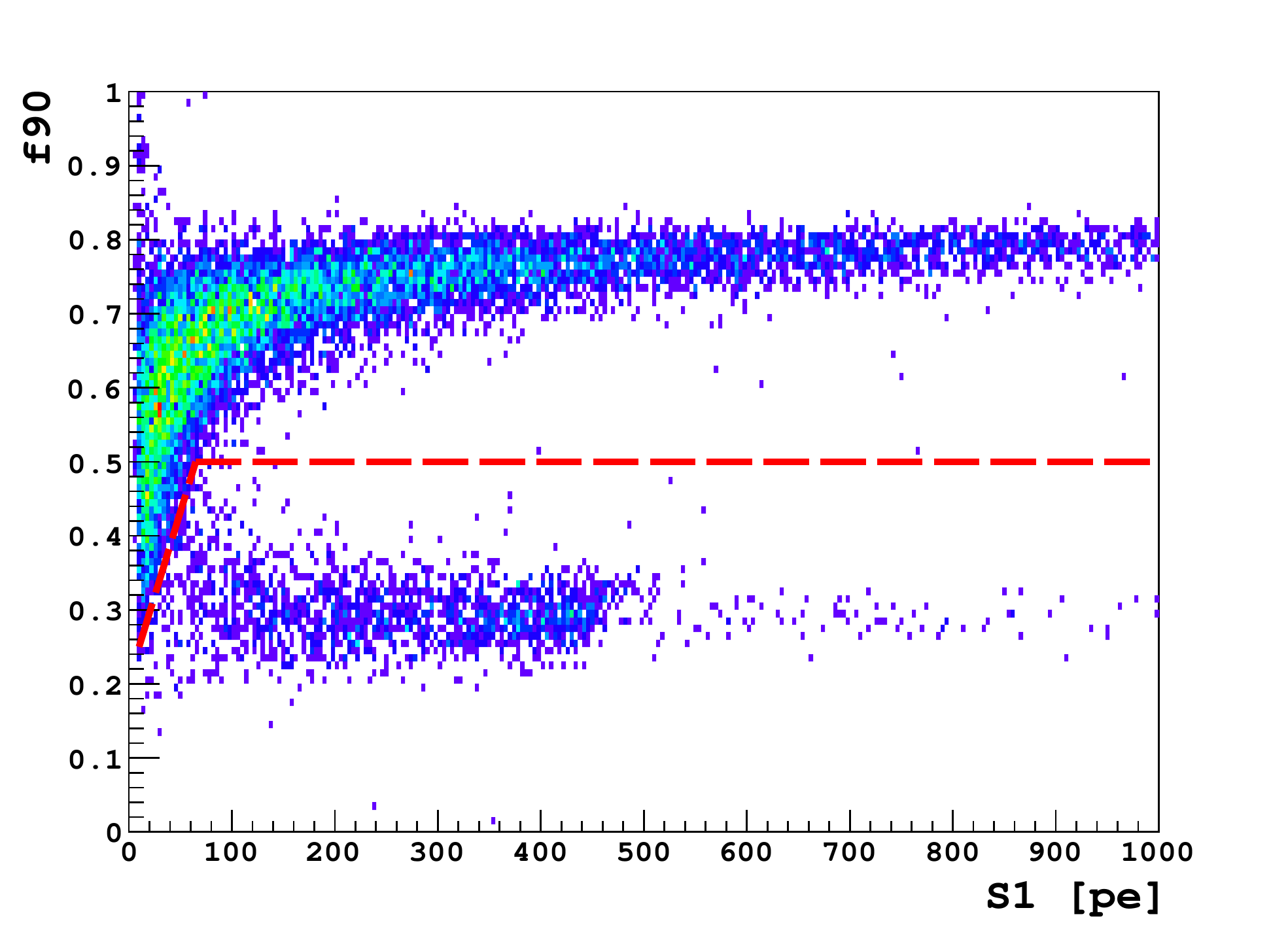}
\caption{f90 vs S1  selection cuts to extract a pure sample of nuclear recoils  from the AmBe calibration data. Same cuts were applied to the G4DS simulation of the  AmBe source. See the text for more details.} 
\label{fig:nspectra}
\end{center}
\end{figure}

Generally, $L_{eff}$ is  defined with respect to the electron recoil energy which is  proportional to the deposited energy. It is usually  extracted at null field, where all electrons and ions are assumed to recombine.  Despite the fact that DarkSide-50 operated at 200 V/cm,  accurate tuning of the recombination probability, as described above, allows one to constrain the $L_{eff}$. 

\begin{figure}[h]
\begin{center}
\includegraphics[width=.70\textwidth]{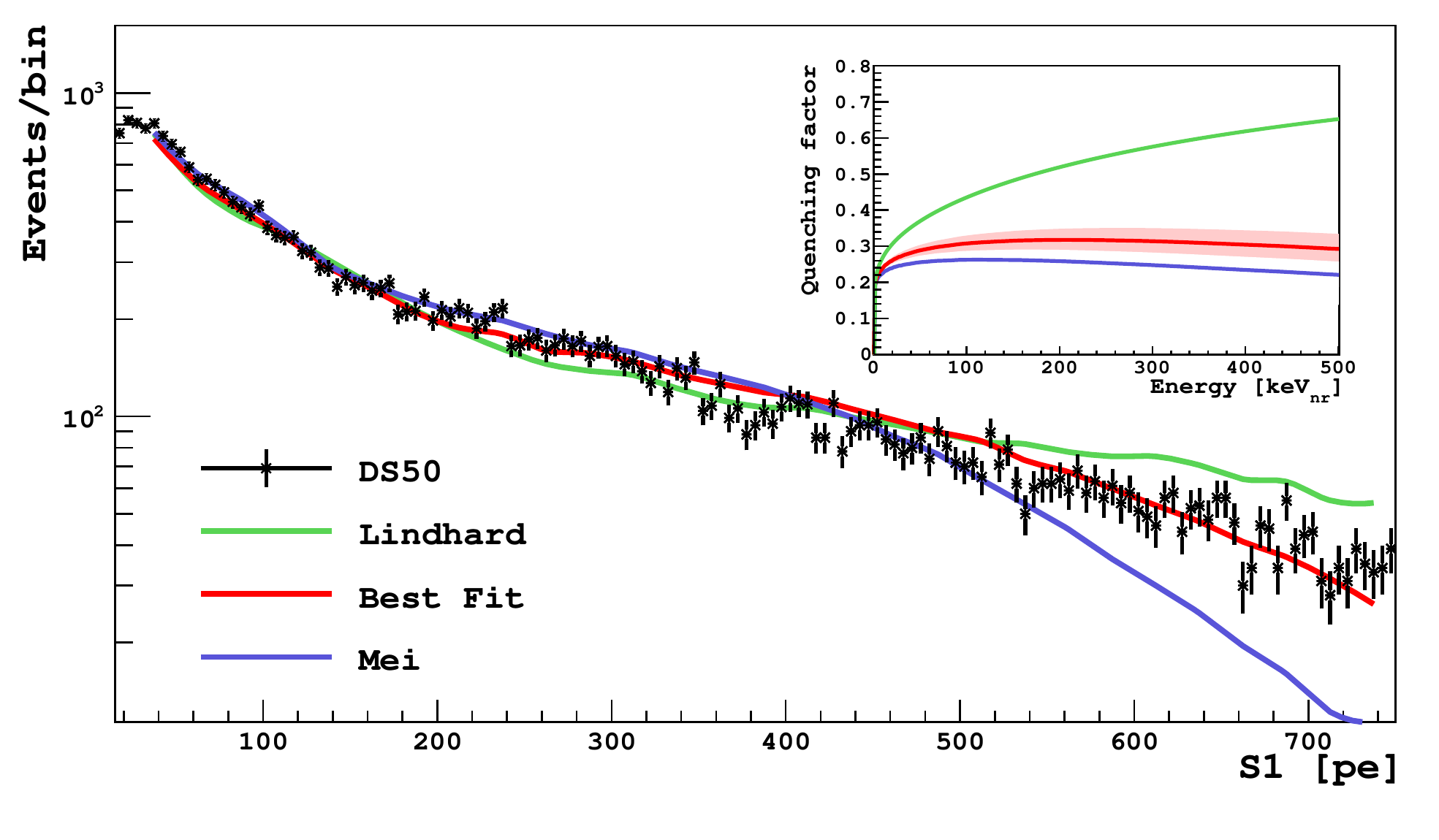}
\caption{Data vs simulation comparison assuming the  Mei and Lindhard models, and with the best fit of the $k_B$ parameter in the modified Mei model. } 
\label{fig:quenching}
\end{center}
\end{figure}

Calibration data using a 10~Bq AmBe neutron source are used to select a sample of nuclear recoils, with a known energy distribution. AmBe emits neutrons in association with a 4.4~MeV gamma ( BR of 65\% ) and coincident 4.4~MeV and 3.2~MeV gammas ( BR of 3\%). These events can be tagged by looking at the coincidence between the liquid scintillator detector and the TPC. The nuclear recoil data was further cleaned from  low energy electron recoil signals in the TPC  by applying  a cut on $f90$. The cut, shown in Figure \ref{fig:nspectra},  was also applied to the simulated data set.

The comparison between DarkSide-50  data and  G4DS with the Lindhard and the Mei models are shown in  Figure \ref{fig:quenching}. Both models are in tension with respect to data, as the Mei model underestimates the quenching factor, while the  Lindhard model overestimates it. Since both models were unable to  reproduce the DarkSide-50 data, a modified Mei model is used as the default quenching model by fitting the single scatter nuclear recoil spectrum obtained from  AmBe neutrons after freeing the $k_B$  parameters. The fit converges for $k_B$ = 4.66$\substack{+0.86 \\ -0.94}$ $\times$ 10$^{-4}$ MeV$^{-1}$ g cm$^{-2}$.  

\begin{figure}[h]
\begin{center}
\includegraphics[width=.75\textwidth]{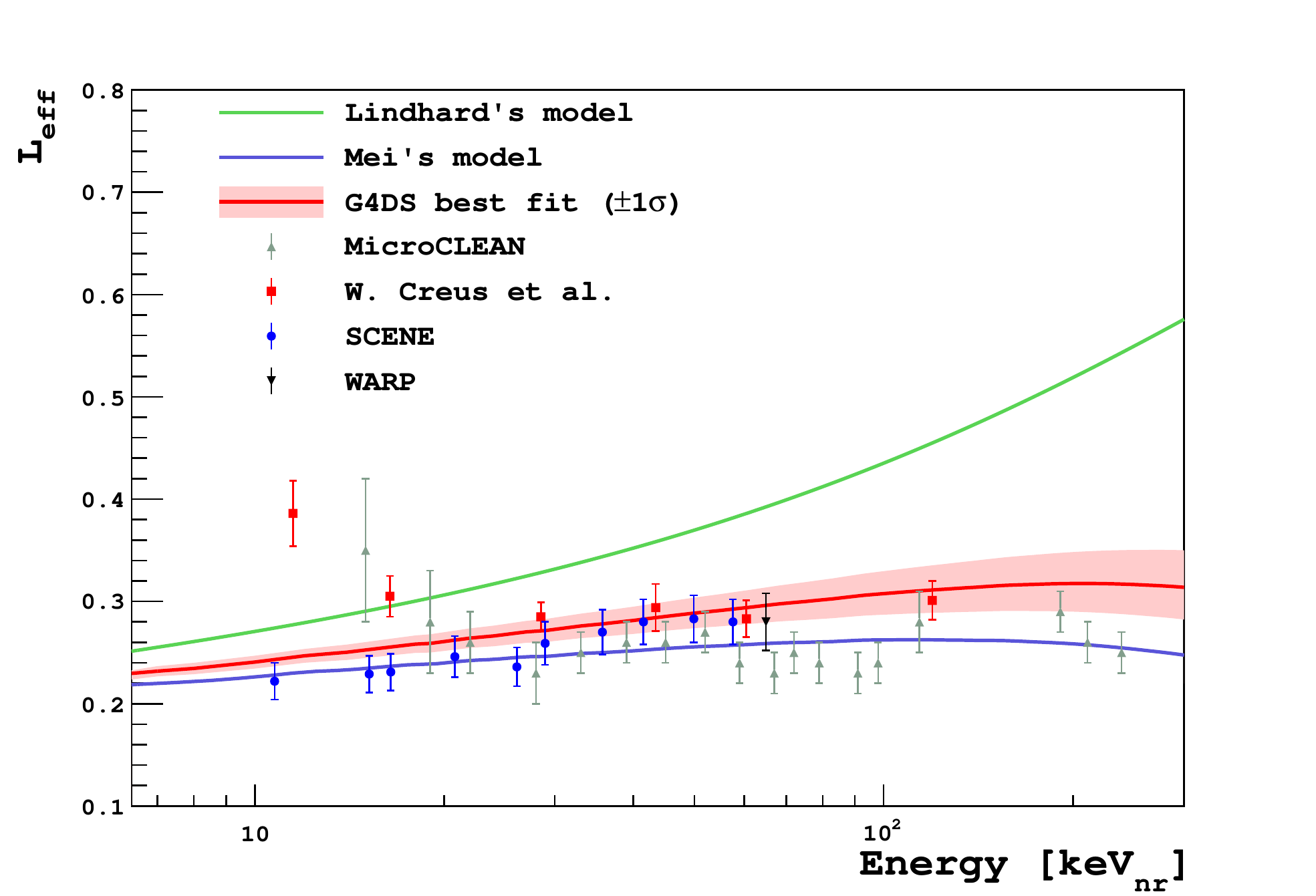}
\caption{Comparison of the Lindhard and Mei models with the G4DS best fit of  the DarkSide-50 data.  SCENE \cite{PhysRevD.88.092006}, MicroCLEAN \cite{PhysRevC.85.065811},  WARP \cite{Brunetti2005265}, and W.~Creus \textit{et al.} \cite{creus:in2p3-01163988} data sets are  also shown.} 
\label{fig:leff}
\end{center}
\end{figure}

The result of the fit is shown also in Figure \ref{fig:leff}, where data sets from the literature are also compared. The so-obtained modified model is in good agreement with the SCENE and WARP measurements, and with  the W.~Creus \textit{et al.} measurements above 20 keV.  A slight variation at $\sim$2~$\sigma$ level arises when  compared to the  MicroCLEAN data set.

\subsection{The S2 response}\label{sec:S2}


The S2 signal provides  information on the number of scatters, the position, and energy of the interacting particle. The simulation of the S2 signal takes into account the dependencies of the optical response on the   radial location of the event, which was discussed in Section \ref{sec:optics}. However, this requires tuning of the electroluminescence yield ($Y_{S2}$). More specifically, the number of photoelectrons detected in  S2  is proportional to the number of ionization electrons which survive recombination and capture on impurities, and the number of emitted  photons per electron is  modelled with a normal distribution centered on $Y_{S2}$ and with a resolution   proportional to $\sqrt{Y_{S2}}$.  The proportionality constant is a free parameter extracted from the calibration. The electroluminescence  light is emitted uniformly  along  the electron drift in the gas pocket, whose duration is of the order of  $\sim$1~$\mu$s, as observed in  the data. We currently use the S1 parameters for the time response of S2; the limited role of  the S2 time profile in the DarkSide-50 analyses  makes the requirements on these parameters less severe.

\begin{figure}[h]
\begin{center}
\includegraphics[width=.49\textwidth]{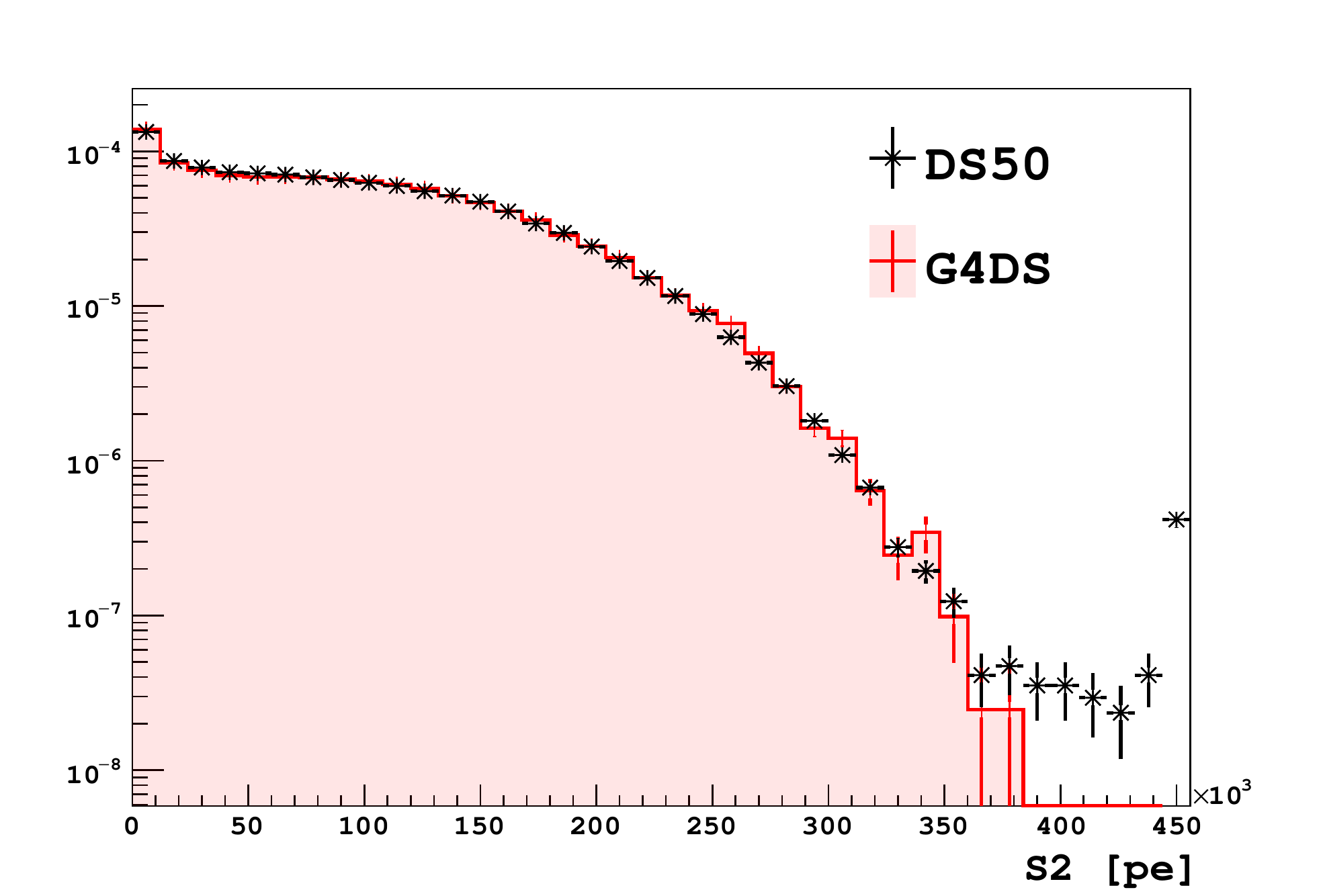}
\includegraphics[width=.49\textwidth]{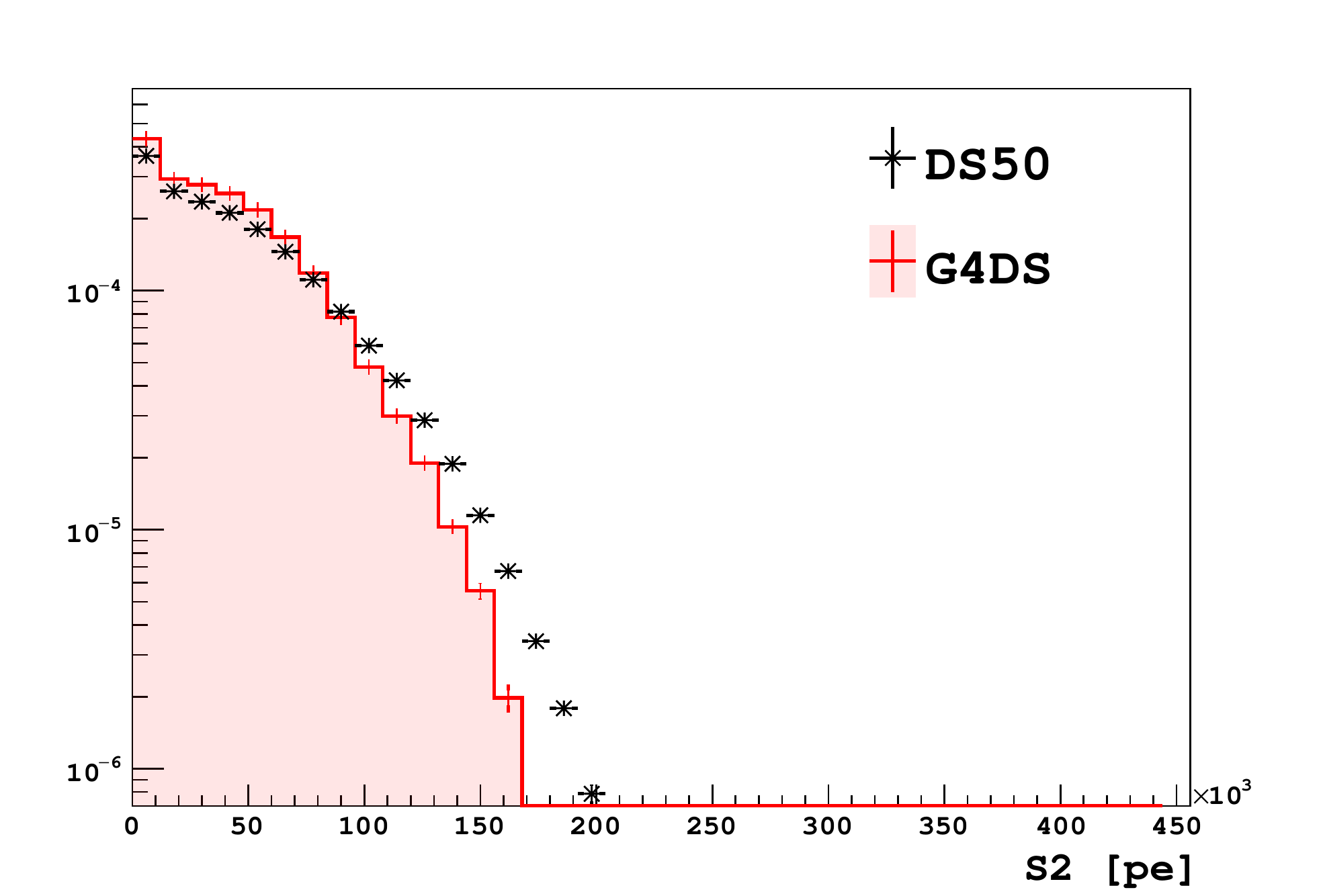}
\caption{Data   and G4DS comparison of the S2 response for  $^{39}$Ar events  in the central column, R$_{xy}$$<$12~cm (left), and in the TPC outer shell, R$_{xy}$$>$12~cm (right).}
\label{fig:s2cal}
\end{center}
\end{figure}

Since S2 of  $^{39}$Ar events saturate the ADC above 5$\times$10$^4$ pe, data were also acquired with a low-gain digitizer (CAEN V1724), which was placed  with a CAEN V1720 during standard data acquisition.  This allowed the detection of the S2 signal with a reduced  amplification (factor of  $\sim$6) which avoided signal saturation.    The radial dependence of S2, discussed in Section \ref{sec:optics}, is  the main source of uncertainty in the S2 tuning. Figure \ref{fig:s2cal} shows the good agreement between data and Monte Carlo for  $^{39}$Ar  events in the central region  (R$_{xy}$$<$12~cm). The mean value for $Y_{S2}$ is 273 photons per extracted electrons, with $\sigma$$\sim1.1\times\sqrt{Y_{S2}}$. The averaged collection efficiency for a photon generated in the gas pocket is 0.163.   The comparison in the  outer shell   ($R_{xy}$$>$12~cm) shows a discrepancy which is connected to   the difficulty in measuring the radial correction at the  TPC edges.

\begin{figure}[h]
\begin{center}
\includegraphics[width=.49\textwidth]{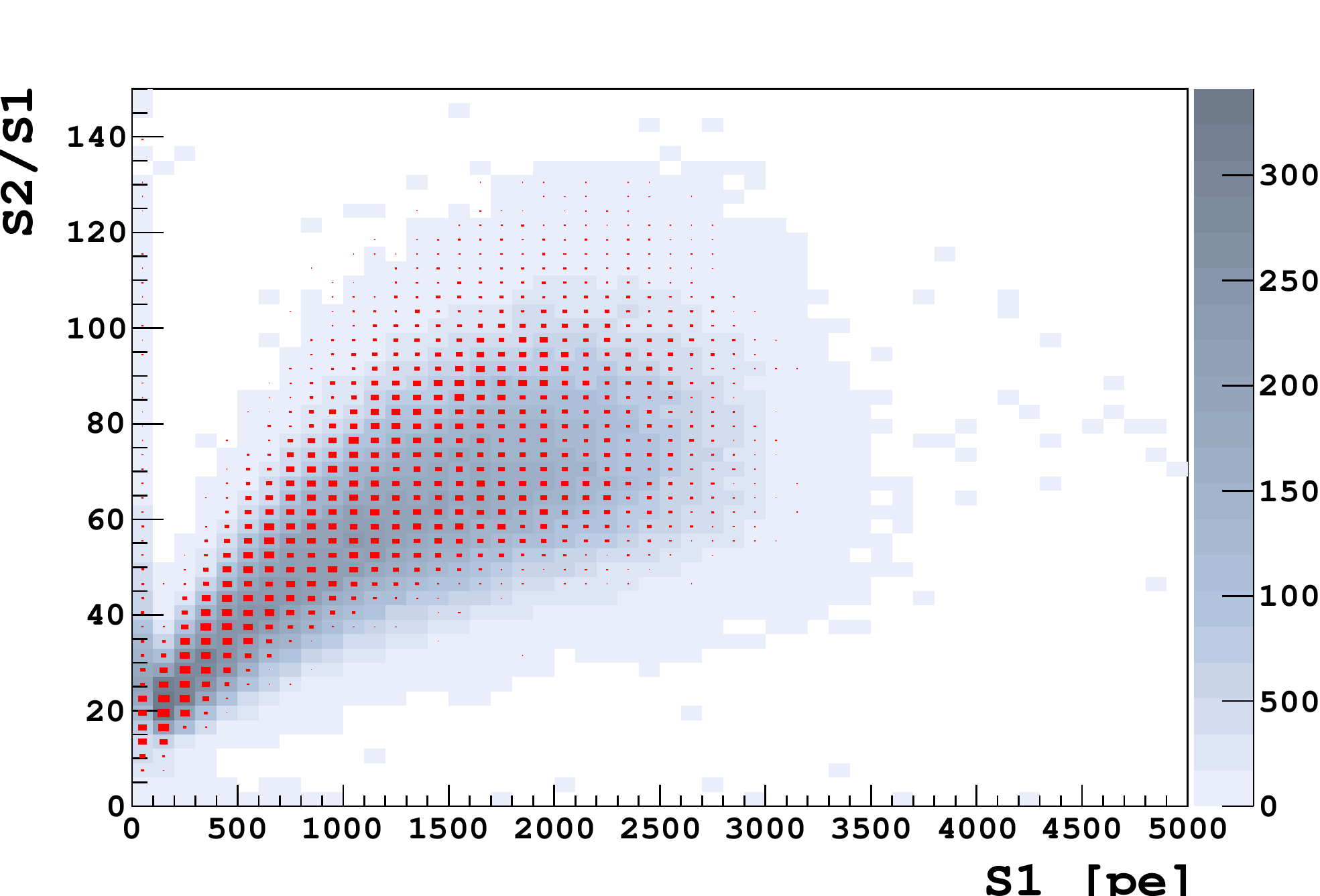}
\includegraphics[width=.49\textwidth]{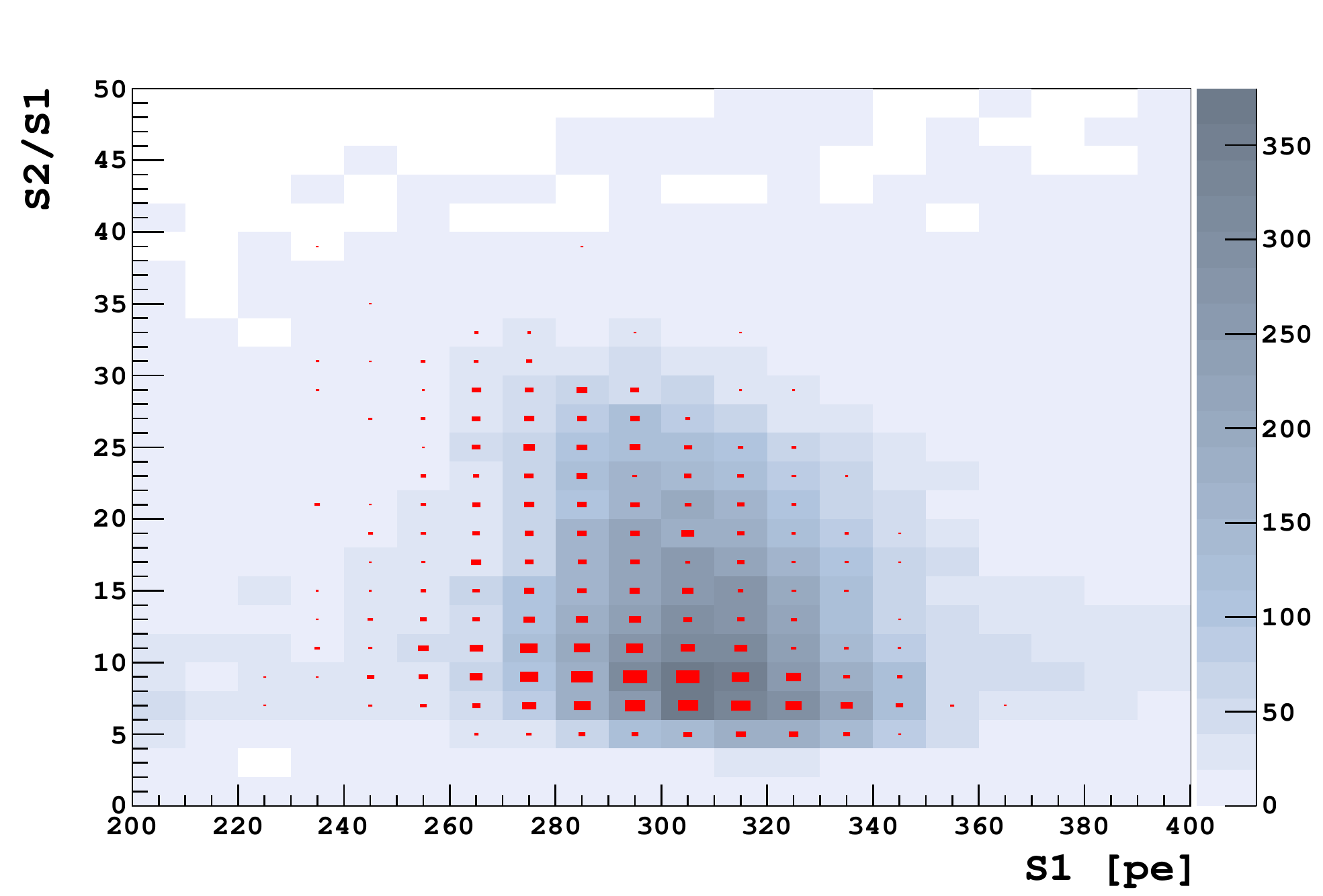}
\caption{Data (grey scale) and G4DS (red boxes) comparison of the S2/S1 ratio versus S1 for  $^{39}$Ar events  in the central column, R$_{xy}$$<$12~cm (left), and  $^{83m}$Kr generated with a spatial distribution derived from real data (right).}
\label{fig:s2s1}
\end{center}
\end{figure}

A second test was performed by comparing  data and simulated $^{83m}$Kr events. However, these are not uniformly distributed in  the LAr volume and have a strongly distorted S2 spectral  shape. To mitigate this effect,  events are simulated from the position distribution of the reconstructed data. Despite the fact that the simulated spectral shape does not fully match the data, the events fall in the same S2 range, and provide an interesting sample to test the correlation between S2 and S1. The observed good agreement is confirmed by  $^{39}$Ar data when selected in the central column, as shown Figure \ref{fig:s2s1}.

\section{Conclusions}
\label{sec:impact}
This work describes the DarkSide-50 simulation framework, G4DS, and its performance when compared with   DarkSide-50 data.  G4DS reproduces energy and time observables at the percent level.  Such a precision was achieved mainly thanks to the careful tuning of the detector optical properties and to  PARIS, the  custom--made effective model describing the physical processes that produce  the S1 and S2 signals.  The  comparisons with DarkSide-50 data presented in this work include  nuclear recoil quenching,  the S1 and S2 response at 200 V/cm drift field, and   pulse shape discrimination.  It is worth mentioning that G4DS is  also  able to precisely reproduce   the detector resolution in S1, demonstrating the robustness of the approach.   

G4DS has been widely exploited in several DarkSide-50 analyses, for example in the identification of the $^{85}$Kr background, in the estimation of the $^{39}$Ar depletion factor in underground argon \cite{Agnes:2015ftt}, and in the prediction of the nuclear recoil background  \cite{Agnes:2014bvk}. Furthermore, its   modular  and flexible structure  allows an extension of G4DS to other detector geometries and physics cases under investigation for future experiments, such as DarkSide-20k \cite{DS20kTDR}, driving the technical design in order to maximize the light yield and the background discrimination power for the WIMP search, and ARGO~\cite{1475-7516-2016-08-017},  investigating the physics potential in the solar neutrino sector at the MeV scale. The recombination law in the PARIS model, in fact, remains valid at the MeV range, though the energy response strongly depends on the local density of ionization. This can differ starkly when going from minimum ionizing tracks, like those induced by neutrino elastic scattering off electrons, to nuclear recoils, as in the WIMP search. The accuracies of the PARIS model, demonstrated at few hundreds of keV with calibration sources, and   of  Geant4 electromagnetic processes, guarantee a comprehensive model, able to range from few keV to the MeV scale.

\acknowledgments
This work was supported by the US NSF (Grants PHY-0919363, PHY-1004072, PHY-1004054, PHY-1242585, PHY-1314483, PHY-1314507 and associated collaborative grants; grants PHY-1211308 and PHY-1455351), the Italian Istituto Nazionale di Fisica Nucleare (INFN), the US DOE (Contract Nos. DE-FG02- 91ER40671 and DE-AC02-07CH11359), the Russian RSF (Grant No 16-12-10369), and the Polish NCN (Grant UMO-2014/15/B/ST2/02561). We thank the staff of the Fermilab Particle Physics, Scientific and Core Computing Divisions for their support. We acknowledge the financial support from the UnivEarthS Labex program of Sorbonne Paris Cit\'e (ANR-10-LABX-0023 and ANR-11- IDEX-0005-02), from S\~ao Paulo Research Foundation (FAPESP) grant  (2016/09084-0), and from Foundation for Polish Science (grant No. TEAM/2016-2/17).

\bibliographystyle{JHEP}
\bibliography{7-biblio.bib}

%

\end{document}